\documentclass[reprint,amsmath,amssymb,aps,preprintnumbers,nofootinbib,showpacs]{revtex4-1}

\usepackage{amssymb}
\usepackage{graphicx}
\usepackage{subfigure}
\usepackage{fullpage}
\usepackage{colortbl}
\usepackage{hyperref}

\begin{document}


\title{Decorrelated Jet Substructure Tagging using Adversarial Neural Networks}

\begin{abstract}
We describe a strategy for constructing a neural network jet substructure tagger which powerfully discriminates boosted decay signals while remaining largely uncorrelated with the jet mass.
This reduces the impact of systematic uncertainties in background modeling while enhancing signal purity, resulting in improved discovery significance relative to existing taggers.
The network is trained using an adversarial strategy, resulting in a tagger that learns to balance classification accuracy with decorrelation.
As a benchmark scenario, we consider the case where large-radius jets originating from a boosted resonance decay are discriminated from a background of nonresonant quark and gluon jets.
We show that in the presence of systematic uncertainties on the background rate, our adversarially-trained, decorrelated tagger considerably outperforms a conventionally trained neural network, despite having a slightly worse signal-background separation power.  We generalize the adversarial training technique to include a parametric dependence on the signal hypothesis, training a single network that provides optimized, interpolatable decorrelated jet tagging across a continuous range of hypothetical resonance masses, after training on discrete choices of the signal mass.

\end{abstract}

\author{Chase Shimmin}
\affiliation{Department of Physics and Astronomy, UC Irvine, Irvine, CA 92627}
\affiliation{Department of Physics, Yale University, New Haven, CT }
\author{Peter Sadowski}
\affiliation{Department of Computer Science, UC Irvine, Irvine, CA 92627}
\author{Pierre Baldi}
\affiliation{Department of Computer Science, UC Irvine, Irvine, CA 92627}
\author{Edison Weik}
\affiliation{Department of Physics and Astronomy, UC Irvine, Irvine, CA 92627}
\author{Edward Goul}
\affiliation{Department of Physics, MIT, Cambridge, MA 02139}
\author{Daniel Whiteson}
\affiliation{Department of Physics and Astronomy, UC Irvine, Irvine, CA 92627}
\author{Andreas S{\o}gaard}
\affiliation{Department of Physics and Astronomy, University of Edinburgh, Edinburgh UK}
\date{\today}

\maketitle

 \section{Introduction}

The enormous center-of-mass energy of the Large Hadron Collider (LHC) enables the production of particles at such extreme velocities that the decay products of even massive particles can become collimated.  Rather than producing distinct deposits of energy in the calorimeter, hadronic decay products of such boosted objects can overlap, creating a single large jet. Distinguishing between jets originating from a single particle (such as a quark or gluon), and those which contain two or three hadronic decay products, is known as jet tagging, and has become an essential component of searches for new physics at the LHC~\cite{Butterworth:2008iy,Adams:2015hiv,Abdesselam:2010pt,Altheimer:2012mn,Altheimer:2013yza}.

However, optimizing the LHC discovery potential requires balancing the competing constraints of signal discrimination and systematic uncertainties.  We consider the case posed in Ref.~\cite{Shimmin:2016vlc} in which a spectrum of jet masses is examined for the presence of a signal-like resonance peak.  The background is dominated by QCD jets, while the hypothetical signal is produced via the hadronic decay of a boosted resonance.

On one hand, there has been intense theoretical work to develop jet substructure tagging tools~\cite{Thaler:2010tr,Larkoski:2013eya} with powerful discrimination between these types of jets.
On the other hand, the processes that produce backgrounds to these searches are often not well understood or are poorly modeled by simulation tools. As a result, experiments in practice rely on the assumption of a smooth background spectrum which can be interpolated under a signal peak from sidebands.
Unfortunately, the jet-tagging quantities may be correlated with jet mass, resulting in a distortion of the background shape~\cite{Dolen:2016kst}, leading to systematic uncertainties which cannot be simply characterized or controlled.
The desire for optimal discrimination and reduced sensitivity to systematic uncertainties are naturally at tension with each other.

One solution, Designing Decorrelated Taggers (DDT)~\cite{Dolen:2016kst}, uses a simple parametric function to construct a modified version of one tagging variable (e.g. $\tau_{21}$), adjusted specifically to avoid distorting the mass spectrum.  This has been shown~\cite{CMS:2016jog} to effectively balance the issues of discrimination and systematic uncertainty for the quantity $\tau_{21}$.

However, a multivariate classifier (such as a neural network) utilizing the full suite of tagging variables will have considerably greater discrimination power than any individual variable, or pair of variables~\cite{Baldi:2016fql}. In principle, the DDT approach could be generalized to handle multiple variables, or even the output of a machine-learning-based combination of these variables, but the more complex and non-linear response will require increasingly complex and non-linear corrections.

In this paper, we incorporate the decorrelation requirement directly into the machine learning strategy by modifying the learning rule to include a constraint which attempts to penalize solutions that distort the background mass spectrum.
The training strategy is adversarial~\cite{schmidhuber_learning_1991,ganin2016,edwards_censoring_2016,NIPS2014_5423}, in which a pair of networks, a classifier and an adversary, are trained simultaneously with different objectives. The classifier is trained in the traditional manner to maximize classification accuracy.  As proposed by Ref.~\cite{Louppe:2016ylz}, the adversary is trained to infer the value of one of the classifier inputs from the classifer response.  In this scheme,  the two networks together perform a constrained optimization which maximizes classification accuracy while minimizing the dependence of the classifier response on the selected input. Here, one network performs jet substruture classification, while the adversary attempts to infer the jet mass solely from the classifier response.

Lastly, we generalize the adversarial decorrelation technique to include the case where both the classifier and its adversary are parameterized by some external quantity, such as a theoretical hypothesis for the mass of a new particle or a field coupling strength.
This is motivated by the fact that resonance searches, such as the one described here, are often performed as scan over a range of potential masses.
Generally the optimal classifier for each hypothesis will differ.
However, the signal simulations used for training can usually only be sampled for a small number of hypotheses values due to the computational expense of producing them.

Networks parameterized in this way~\cite{Cranmer:2015bka,Baldi:2016fzo} can interpolate to provide optimal classification for hypotheses which were not included in the training, allowing sensitivity to be evaluated without generating simulations at those points.
We show that a single adversarially-trained classifier, parameterized in the hypothesis signal mass, remains decorrelated over the range of values upon which it is trained.

\section{Benchmark Data}

\begin{figure}[h!]
\begin{center}
\includegraphics[width=0.3\textwidth]{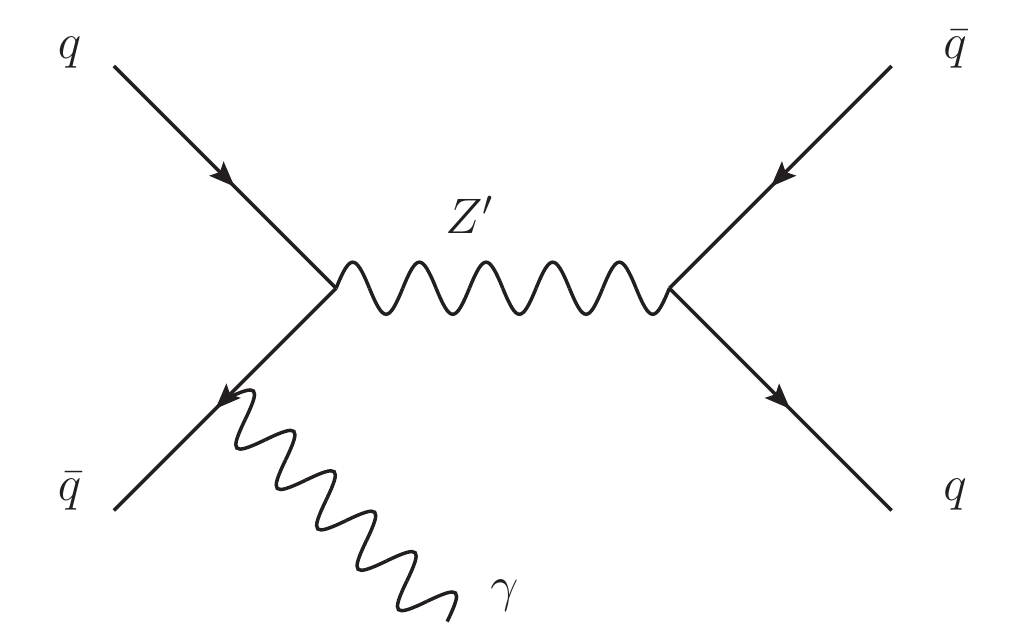}
\caption{Diagram of a hadronically-decaying resonance ($Z'$) produced recoiling against an initial state photon ($\gamma$).}
\label{fig:feyn}
\end{center}
\end{figure}

Simulated samples are used to model the kinematics of the signal and background processes. As a benchmark signal, we use the $Z'$ model from Ref.~\cite{Shimmin:2016vlc}, which produces a hadronically-decaying resonance boosted by its recoil against an initial state photon (Fig.~\ref{fig:feyn}). The same model can be used to study recoil against initial-state gluons or $W$ bosons; we choose the photon channel due to the simpler event topology.

Signal events in which a hypothetical $Z'$ boson decays to quarks are simulated at parton level with {\sc madgraph}5~\cite{Alwall:2014hca} v2.2.3, with {\sc pythia}~\cite{Sjostrand:2006za} v6.4.28 for showering and hadronization, and with {\sc delphes}~\cite{deFavereau:2013fsa} v3.1.2 in the ATLAS-style configuration for primitive detector simulation. The primary background is due to $\gamma+$jets production, which is generated with {\sc sherpa}~\cite{Gleisberg:2008ta} v.2.2.0 requiring one photon and one to three additional hard partons.  

The measurement of jet masses is sensitive to the presence of additional in-time $pp$ interactions, referred to as {\it pile-up} events.  We overlay such interactions in the simulation chain, with an average number of interactions per event of $\left<\mu\right>=15$, which is comparable to the level observed in ATLAS 2015 data, with the LHC delivering collisions at a 25ns bunch crossing interval.

The impact of pile-up events on jet reconstruction can be mitigated using several techniques. First, we employ a jet-area-based pileup subtraction on narrow-radius jets, as implemented by {\sc fastjet}~\cite{Cacciari:2011ma}.  Additionally, when reconstructing large-radius jets, we apply a jet-trimming algorithm~\cite{Krohn:2009th} which is designed to remove pileup, while preserving the two-pronged jet substructure characteristic of boson decay.  Jets are trimmed by reclustering into $k_{\textrm{T}}$ subjets, with $R_{\textrm{trim}}=0.2$, and dropping subjets with less than 3\% of the original jet $p_{\textrm{T}}$. 

As the angular separation of the quarks may be quite small in the case of a high-$p_{\textrm{T}}$ $Z'$, we  reconstruct a single large-radius jet with distance parameter $R=1.0$.   To reflect the thresholds imposed by the ATLAS trigger, we require $p_{\textrm{T}}^{\gamma}>150$ GeV and $p_{\textrm{T}}^{\textrm{jet}}>150$ GeV. In the case of multiple large-$R$ jets, the one with greatest $p_{\textrm{T}}$ is selected.

For the large-radius jets, we calculate various jet substructure variables such as the $N$-subjettiness ratio $\tau_{21}$~\cite{Thaler:2010tr,Thaler:2011gf}, and the Energy Correlation Functions~\cite{Larkoski:2013eya,Larkoski:2014gra}. Recent studies have shown that deep neural networks applied to lower-level calorimeter information can match the performance of several of these higher-level variables in combination~\cite{Baldi:2016fql}, but these higher-level variables capture most of the discriminative information and are theoretically well understood.

\begin{figure}[h!]
\begin{center}
\includegraphics[width=0.2\textwidth]{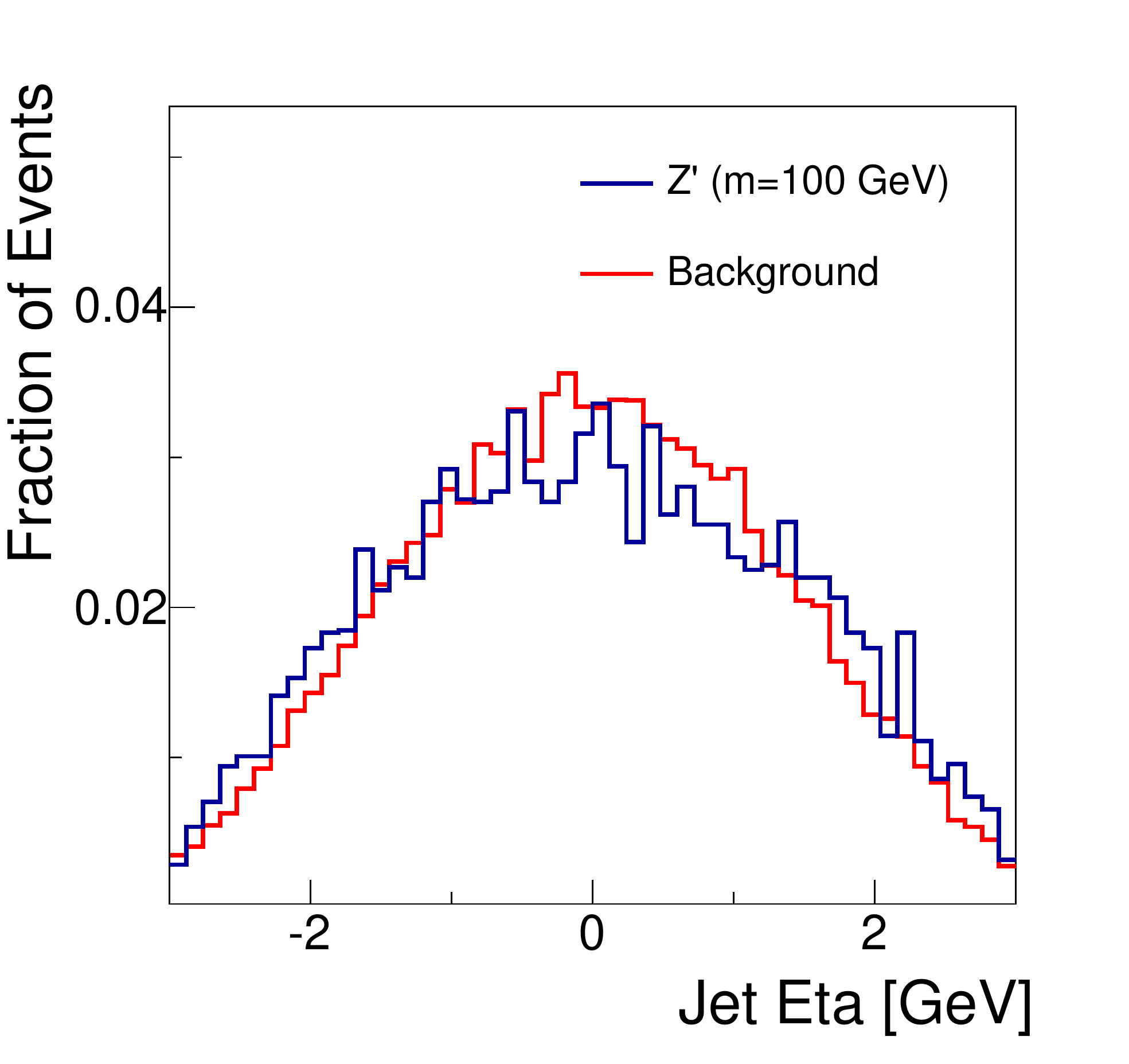}
\includegraphics[width=0.2\textwidth]{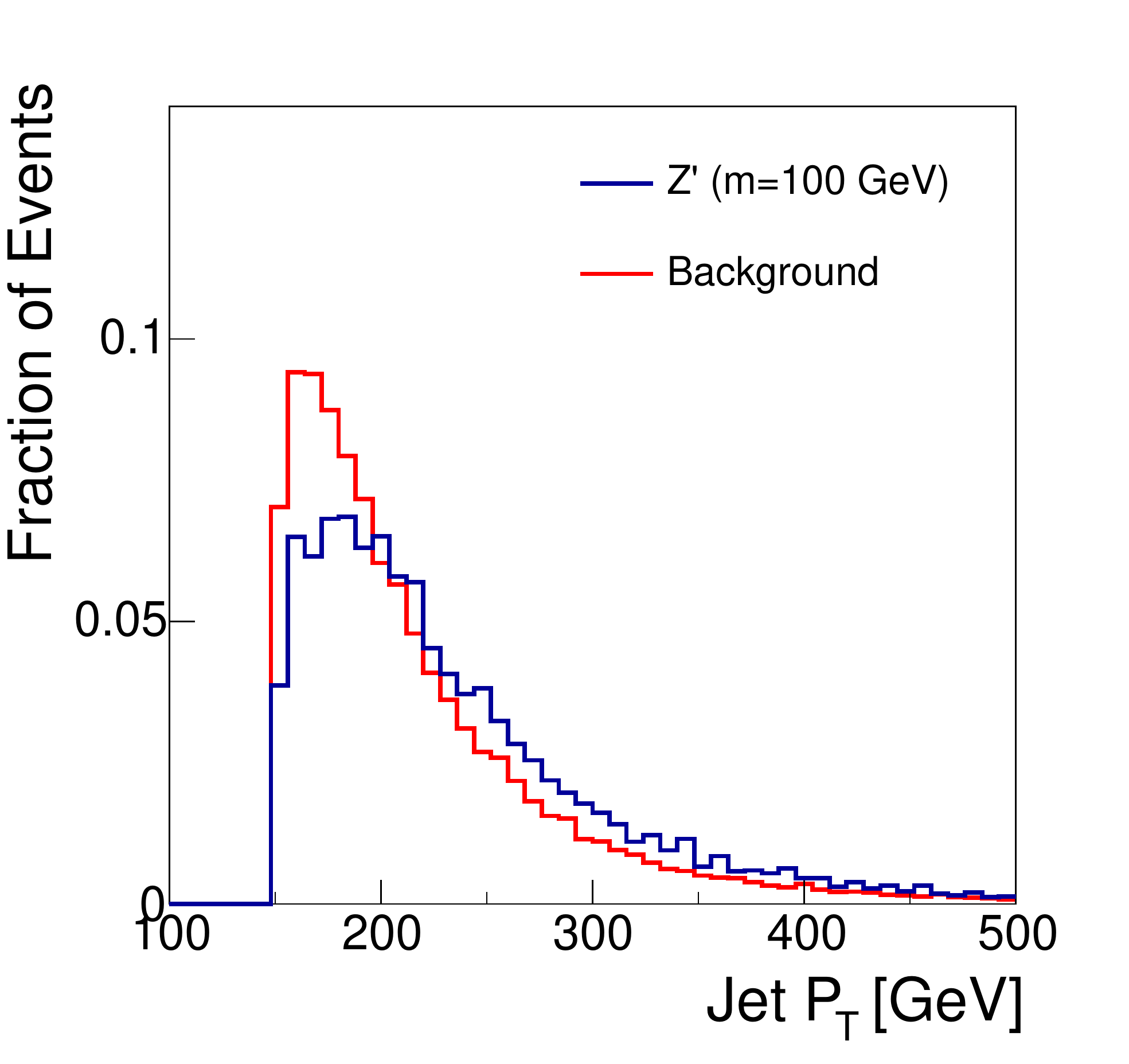}
\includegraphics[width=0.2\textwidth]{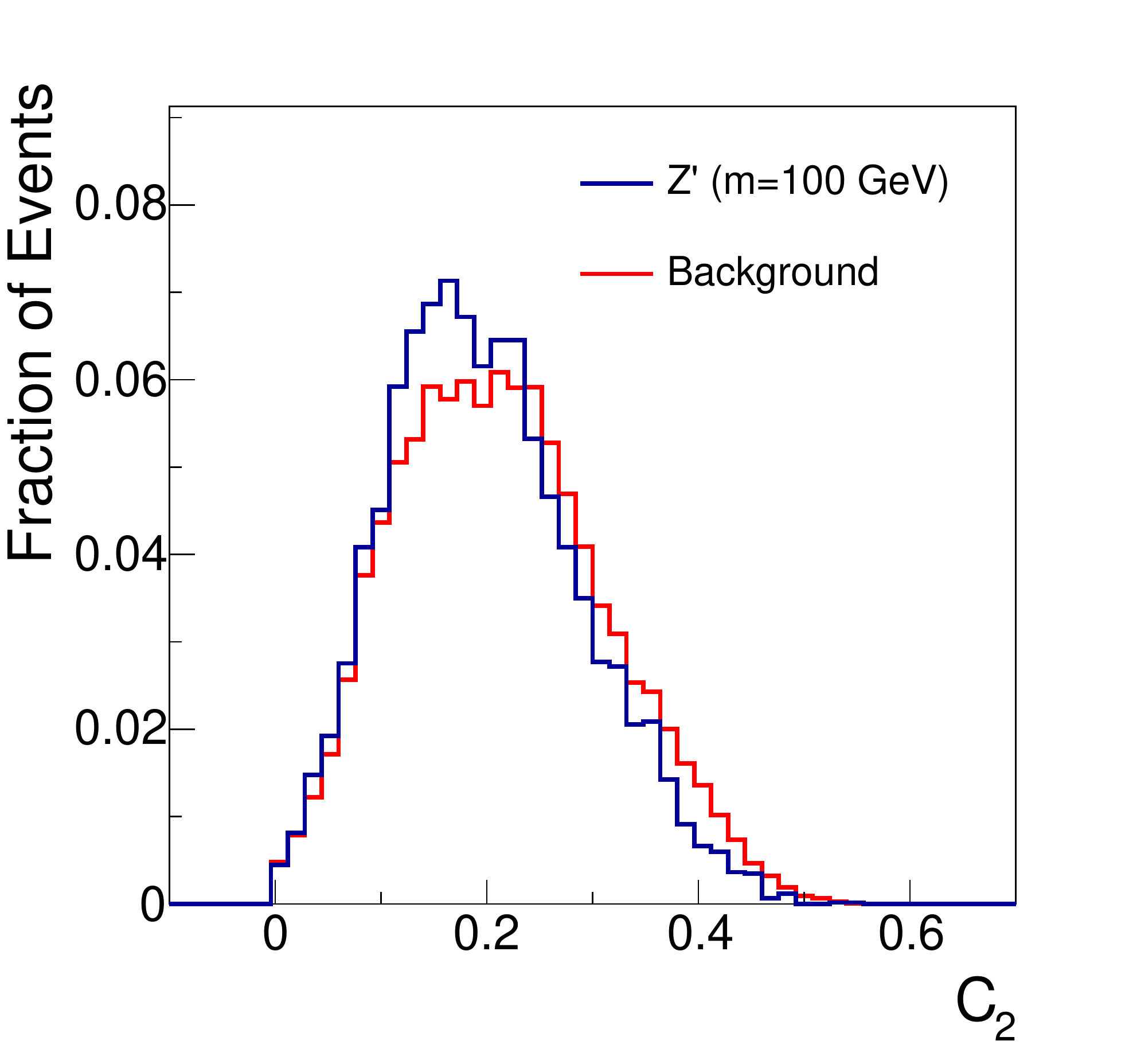}
\includegraphics[width=0.2\textwidth]{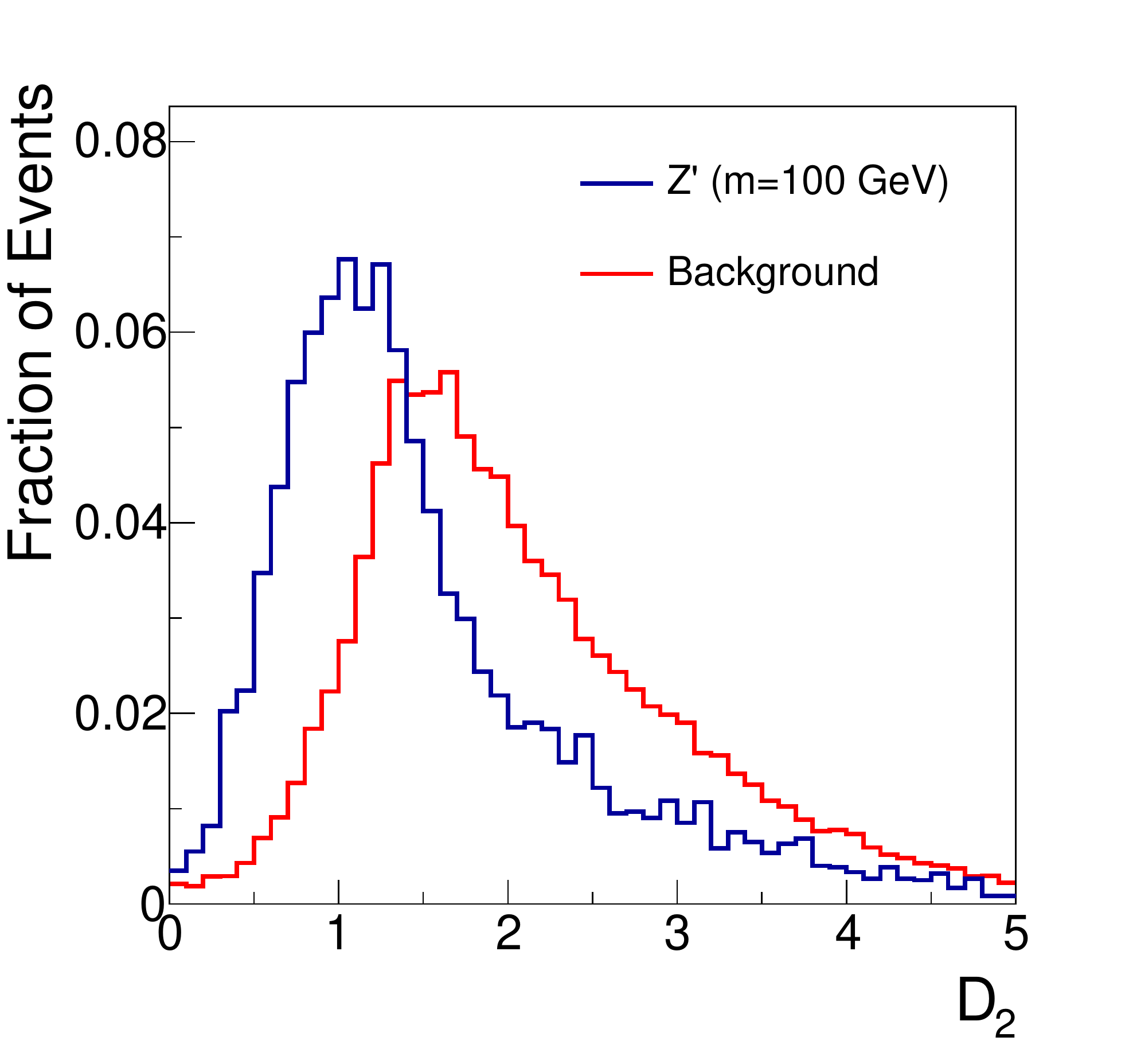}
\includegraphics[width=0.2\textwidth]{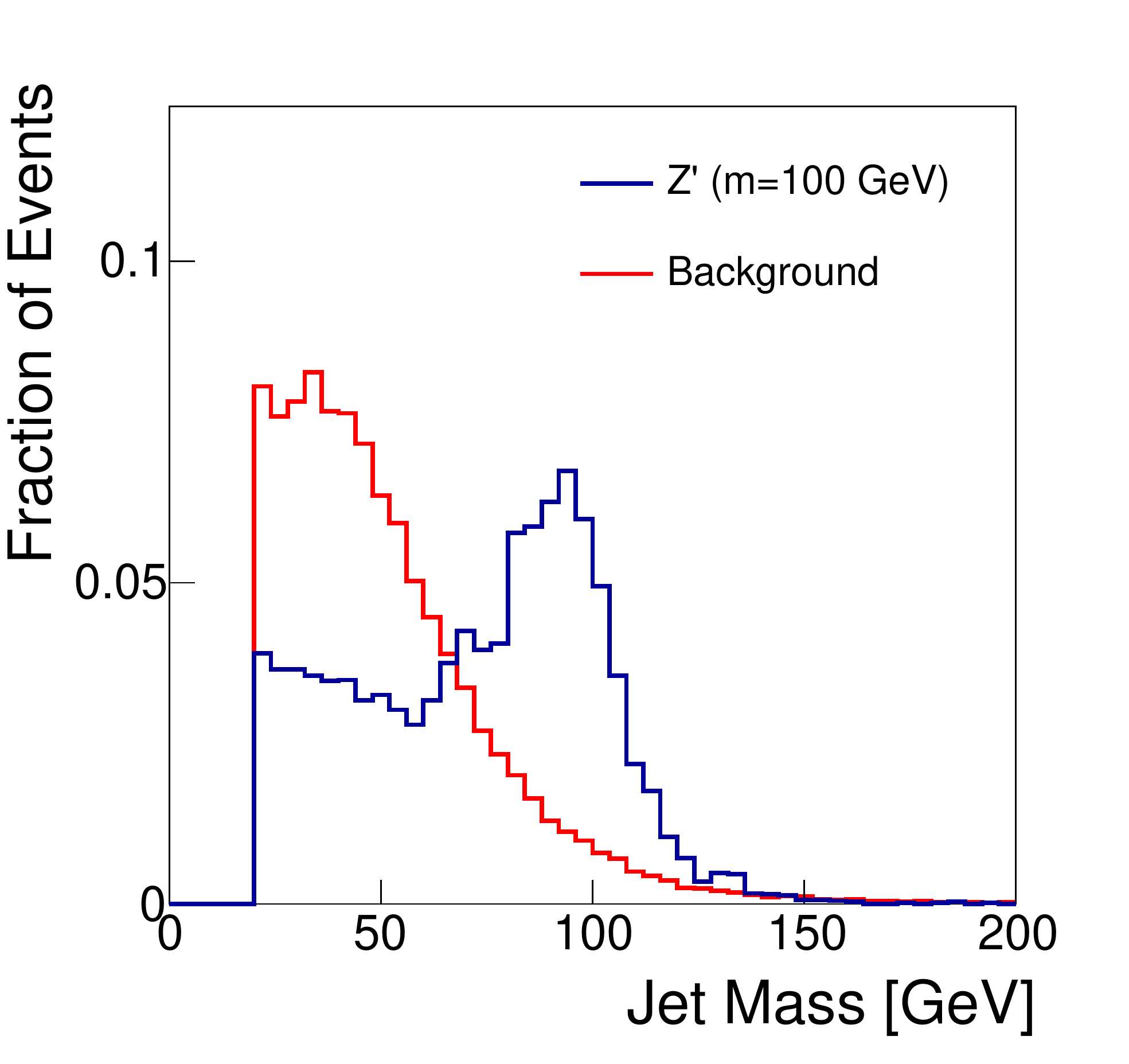}
\includegraphics[width=0.2\textwidth]{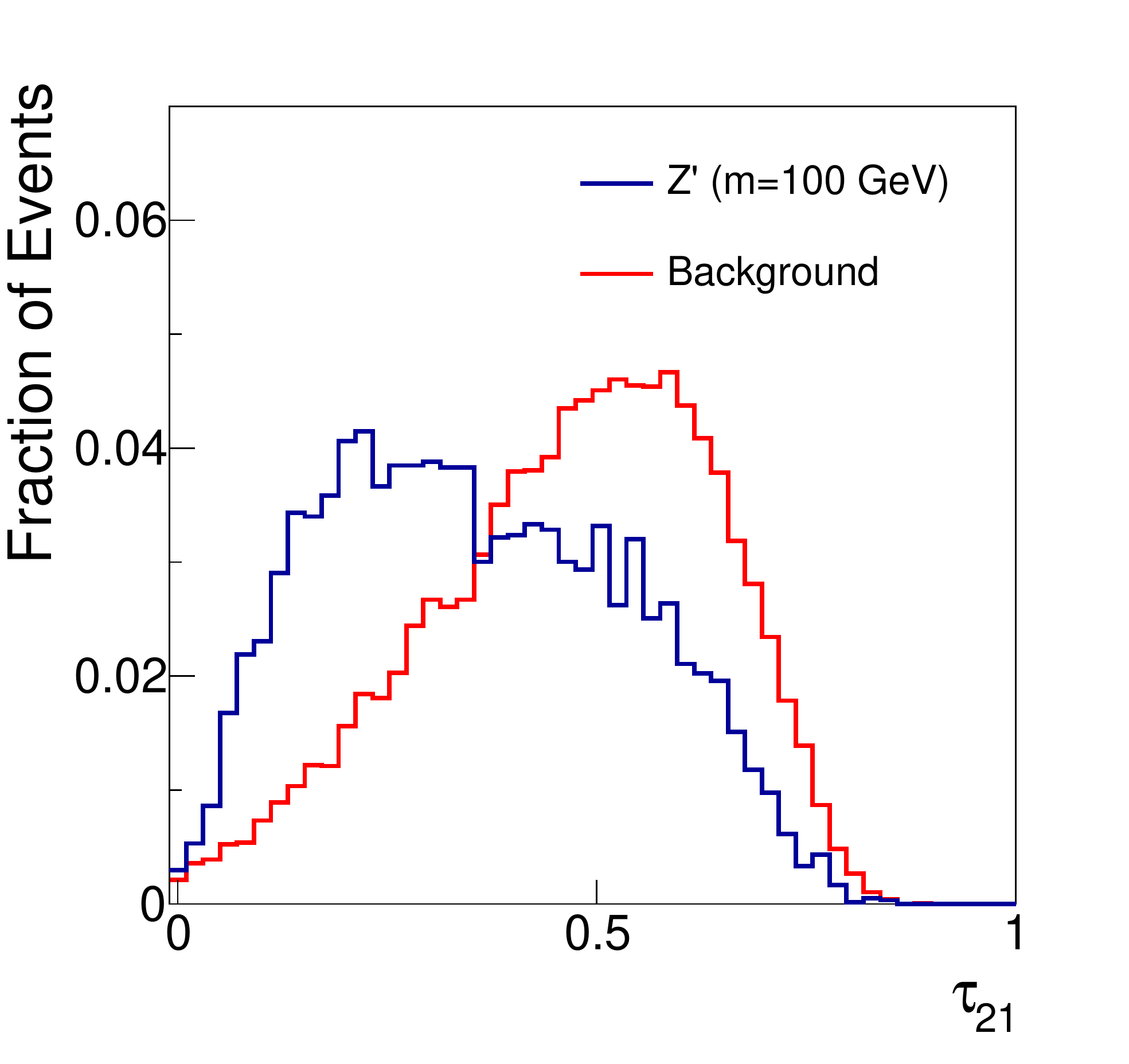}
\caption{Distributions of jet variables in simulated $Z'+\gamma$ signal events, with $m_{Z'}=100$ GeV, as well as $\gamma+$jet background events. From top left to bottom right are shown the jet pseudorapidity, transverse momentum, energy correlation variables $C_2$ and $D_2$~\cite{Larkoski:2013eya}, jet invariant mass, and N-subjettiness($\tau_{21}$)~\cite{Thaler:2010tr}. There are five additional input variables described in the text (not shown).}
\label{fig:var}
\end{center}
\end{figure}

Distributions of the various kinematic quantities for jets selected in signal and background processes are shown in Fig.~\ref{fig:var}. The neural networks described below use eleven variables:
\begin{itemize}
\item Jet pseudo-rapidity, azimuthal angle, transverse momentum, and invariant mass;
\item Jet energy correlation variables, $C_2$ and $D_2$~\cite{Larkoski:2013eya};
\item Jet N-subjettiness ($\tau_{21}$)~\cite{Thaler:2010tr}; and
\item Photon energy, pseudo-rapidity, azimuthal angle, transverse momentum.
\end{itemize}

For comparison with Ref.~\cite{Dolen:2016kst}, we additionally apply the DDT procedure to produce a modified variable, $\tau'_{21}$, which has reduced correlation with jet mass. However, no simple linear relationship was seen between the profile of $\tau_{21}$ and the jet mass, and a linear correction does not remove the dependence; this may be due to the application of jet trimming, which differs from the treatment in Ref.~\cite{Dolen:2016kst}. To provide a fair comparison, we extend the DDT-style approach to use a second-order correction, producing a variable $\tau''_{21}$, which demonstrates reasonable indepenence from the jet mass (Fig.~\ref{fig:profs}).

\section{Neural Networks}

The strategy outlined in Ref.~\cite{Louppe:2016ylz} describes how to train a classifier which is uncorrelated with a nuisance parameter. Here, we apply this strategy to the closely-related problem of decorrelating the classifier with respect to the jet invariant mass, as the nuisance parameter is not well defined; further discussion of this issue is found below in Sec.~\ref{sec:stats}. In Sec.~\ref{sec:param}, we extend this strategy to a problem requiring a parameterized solution.

\begin{figure}[h!]
\begin{center}
\includegraphics[width=0.45\textwidth]{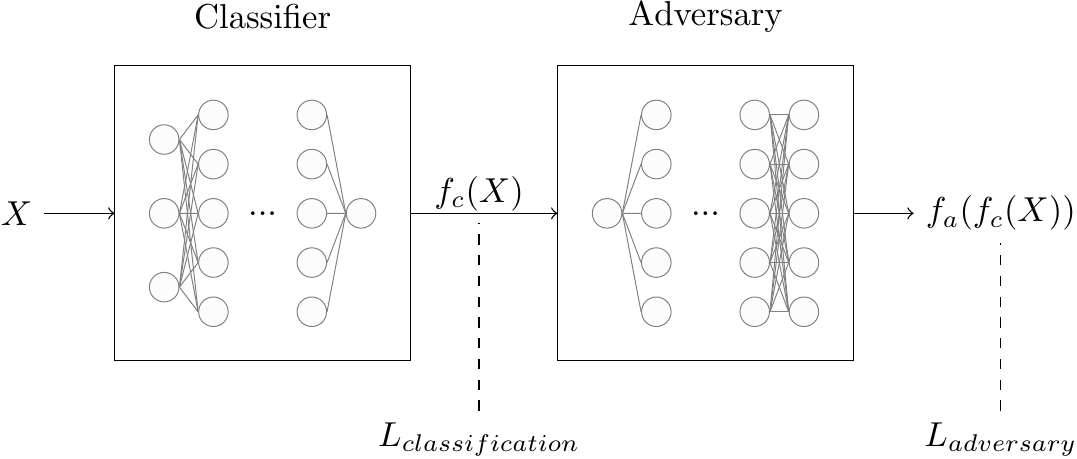}
\caption{ Architecture of the neural networks in the adversarial training strategy. The classifying network distinguishes signal from background using the eleven variables ($X$) described in the text. The adversarial network attempts to predict the invariant mass using only the output of the classifier, $f_{c}(X)$; note that the adversary has multiple binary classification outputs, corresponding to bins in jet invariant mass, rather than a single regression output.}
\label{fig:network}
\end{center}
\end{figure}

Two neural networks --- a jet classifier and an adversary --- constitute two distinct segments of the feedforward architecture shown in Fig.~\ref{fig:network}. The loss of the tagger is defined as

\[ L_\textrm{tagger} = L_{\textrm{classification}} - \lambda L_{\textrm{adversary}}, \]

\noindent
where $\lambda$ is a positive constant, and $L_{\textrm{classification}}$ and $L_{\textrm{adversary}}$ are the standard classification-error loss functions for each segment. The two neural networks are trained concurrently; the tagger's objective is to minimize $L_{\textrm{tagger}}$, while adversary minimizes only $L_{\textrm{adversary}}$. The hyperparameter $\lambda$ represents a tradeoff between the two objective terms; we found that a value of $\lambda=100$ was a good tradeoff for our task, but in general this hyperparameter can be optimized like any other.

The classifier network in this experiment consisted of eleven input features, three fully-connected hidden layers each with 300 nodes having hyperbolic tangent activation functions, and a single logistic output node with the binomial cross-entropy classification objective. The adversarial network consisted of a single input, 50 nodes with hyperbolic tangent activation functions, and a softmax output layer with 10 classes corresponding to binned values of the jet invariant mass (each bin representing one decile of the background), and the multi-class cross-entropy classification objective. 

Because the adversary is challenged with adapting to an ever-changing input as the classifier is trained, and also because its task is relatively easy, two strategies were used to train the adversary faster than the classifier. First, the adversary was given a head start at the beginning of training with 100 updates while the classifier was fixed. Second, the adversary was trained with a larger learning rate of $1.0$ compared to $10^{-3}$ for the tagger objective.

The data set used for experiments was divided into training (80\%), validation (10\%, used for hyperparameter tuning), and testing (10\%) subsets. Each classifier input feature was log-scaled if the empirical skew estimate was greater than 1.0, then standardized to zero mean and unit variance. Model parameters were initialized from a scaled normal distribution~\cite{glorot_2010}.

Training was performed using stochastic gradient descent, applied to mini-batches of 100 examples from each class. During training, the event weights were scaled so that the average weight for each class was 1.0.
However, in the adversarial loss function $L_{\textrm{adversary}}$, the signal events were given zero weight, rendering them invisible to the adversary.

Updates were made using a training momentum term of 0.5; the learning rate decayed by a factor of $10^{-5}$ after each update. Training was stopped after 100 epochs, where an epoch was defined as a single pass through the background samples ($\approx 400$k training events). Models were implemented in {\sc Keras}~\cite{chollet_keras_2015} and {\sc Theano}~\cite{theano2016}, and hyperparameters were optimized on a cluster of Nvidia Titan Black processors.

\section{Performance}

\begin{figure}[h!]
\begin{center}
\includegraphics[width=0.45\textwidth]{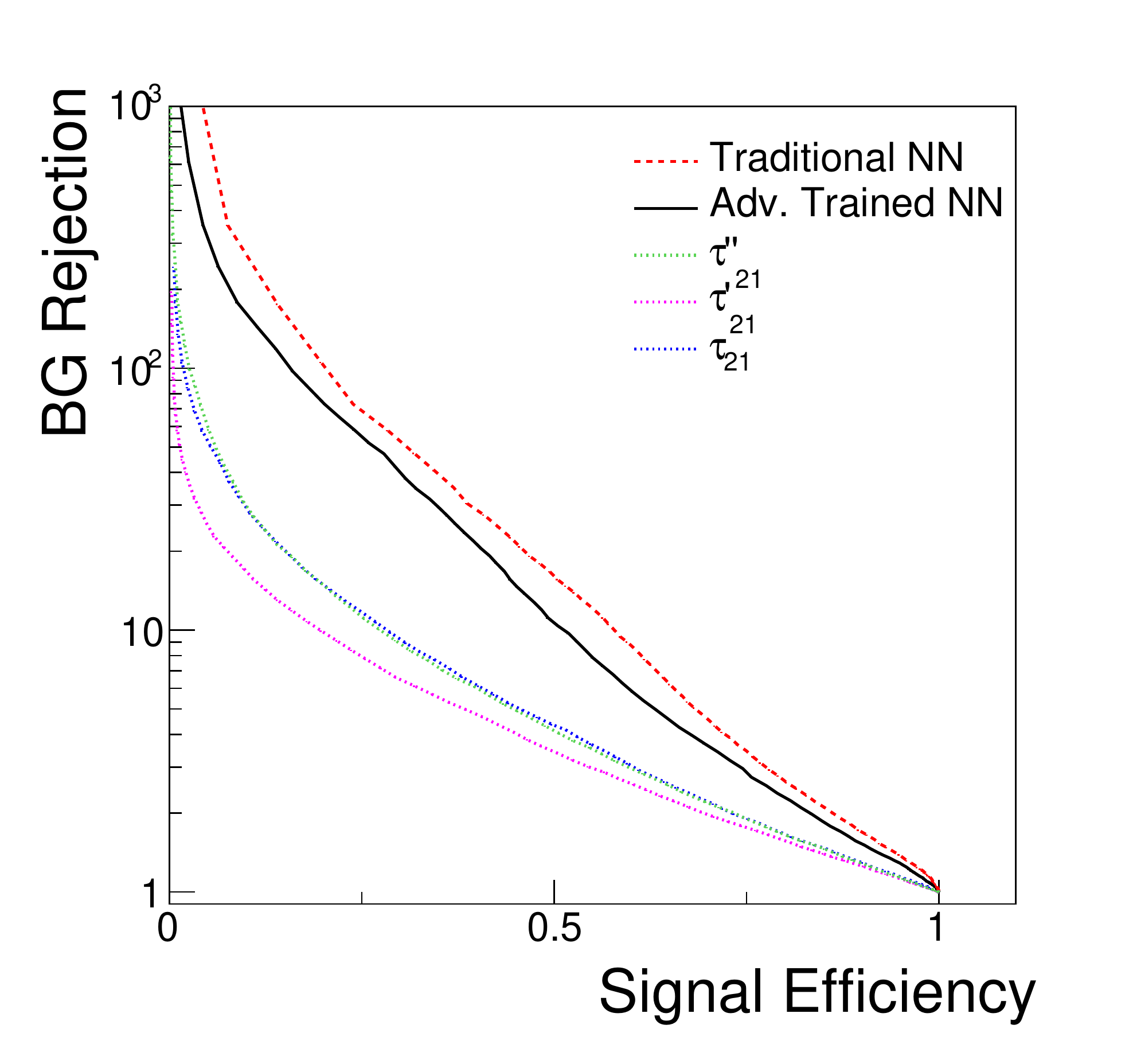}
\caption{ Signal efficiency and background rejection ($1/$efficiency) for varying thresholds on the outputs of several jet-tagging discriminants: traditional networks trained to optimize classification,  networks trained with an adversarial strategy to optimize classification while minimizing impact on jet mass, the unmodified $\tau_{21}$, and the two DDT-modified variables $\tau'_{21}$, and $\tau''_{21}$. The signal samples have $m_{Z'}=100$ GeV for this example. Generalization to other masses is shown in Sec.~\ref{sec:param}.}
\label{fig:roc}
\end{center}
\end{figure}

We compare the discrimination power of five candidate classifiers: the NN trained without an adversary, the adversarially-trained NN, the unmodified $\tau_{21}$, and the two DDT-modified variables $\tau'_{21}$, and $\tau''_{21}$.
The performance can be characterized by measuring the signal efficiency and background rejection of various thresholds on these discriminators (Fig.~\ref{fig:roc}).

The variable $\tau'_{21}$, which is modified to reduce correlation with the mass, results in a modest decrease in its classification power relative to the unmodified $\tau_{21}$ at $m_{Z'}=100$ GeV, though note that these effects are mass-dependent for both $\tau'_{21}$ and $\tau''_{21}$.
Similarly, the adversarial network does not match the discrimination power of the traditional classification network, due to the additional constraint imposed in its optimization. However, both NNs are clearly able to take advantage of the combined power of the substructure variables, and offer a large improvement in background rejection for similar signal efficiencies compared to classification based on $\tau_{21}$ alone.

\begin{figure}[h!]
\begin{center}
\includegraphics[width=0.23\textwidth]{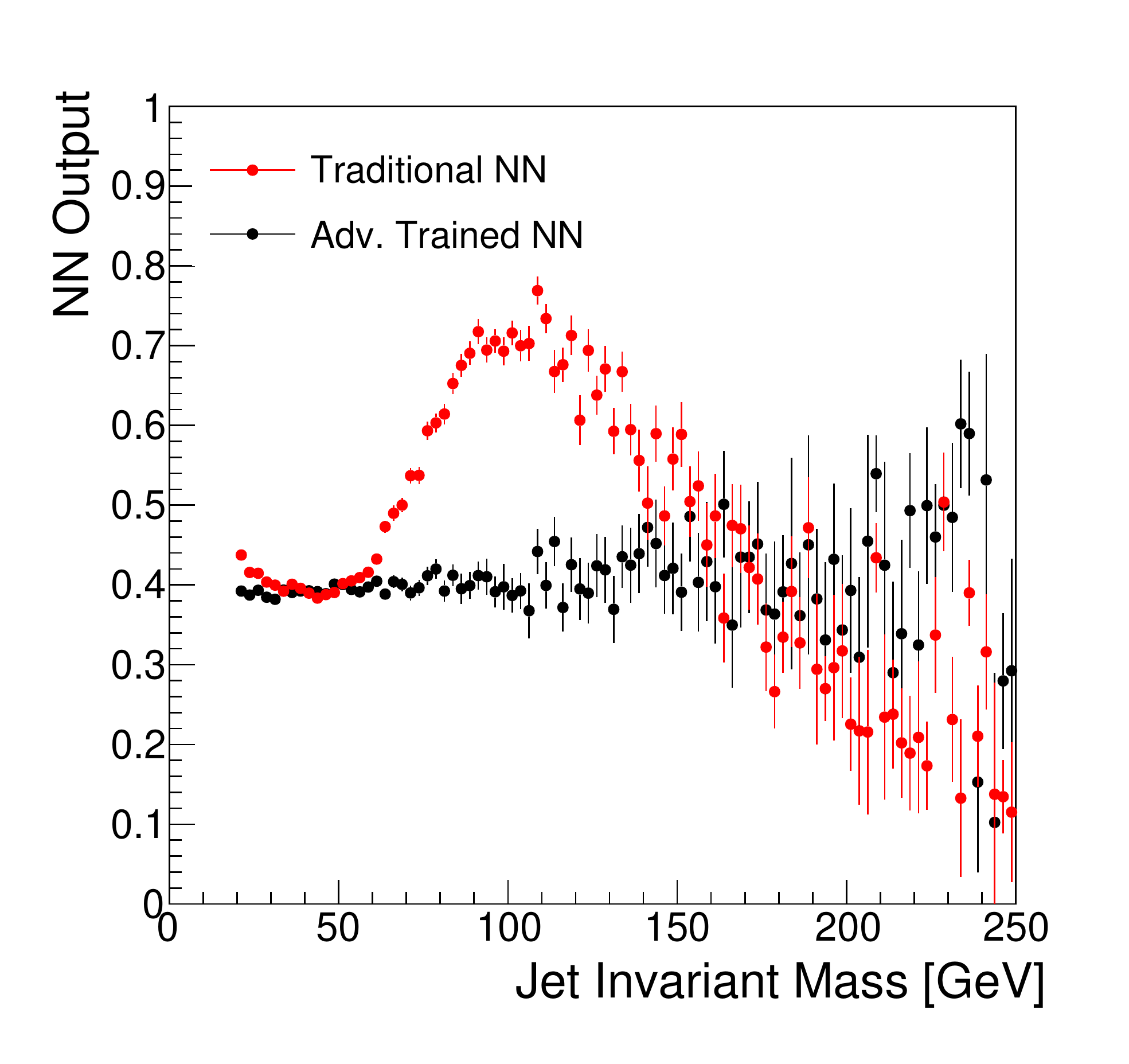}
\includegraphics[width=0.23\textwidth]{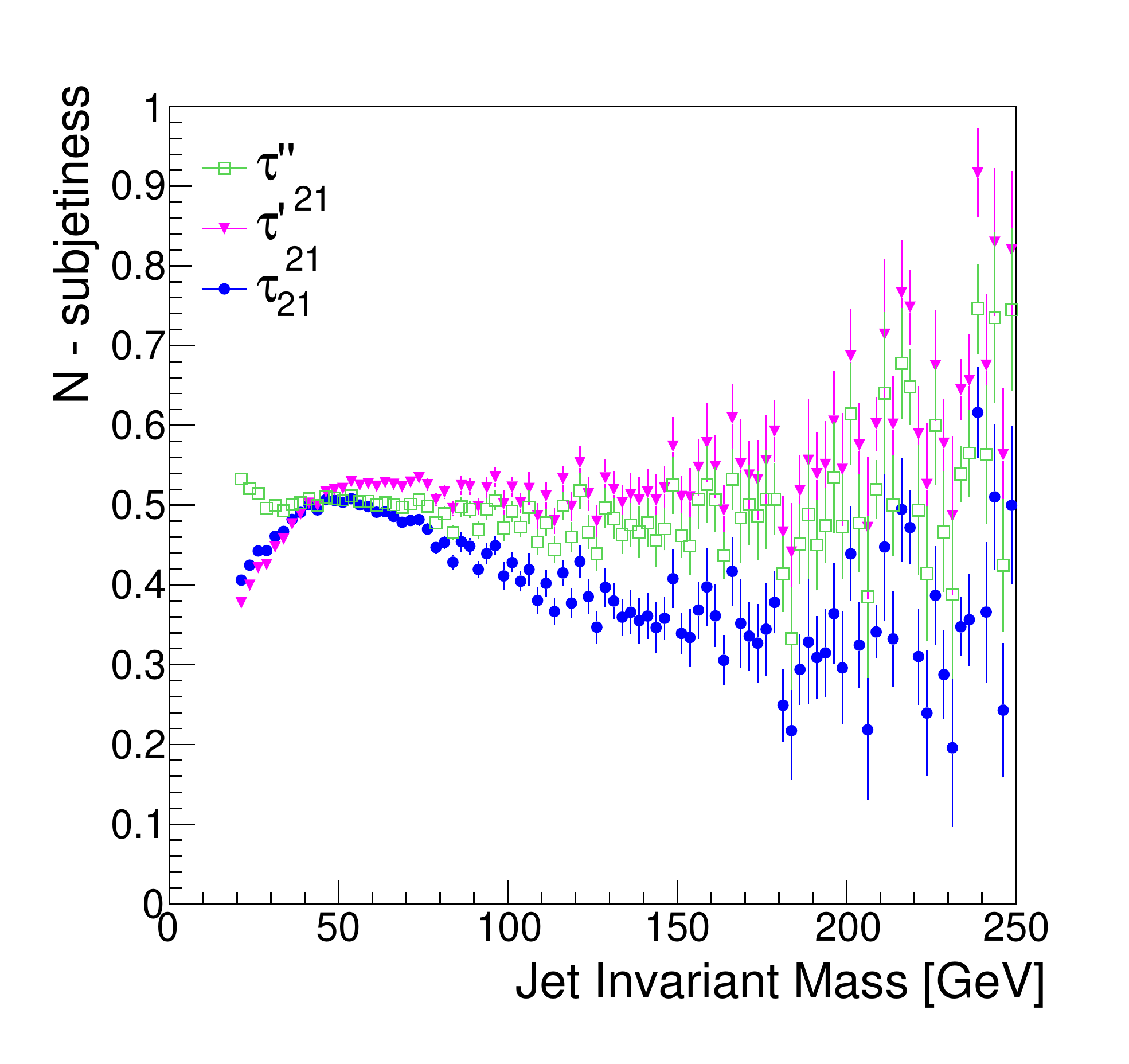}
\includegraphics[width=0.23\textwidth]{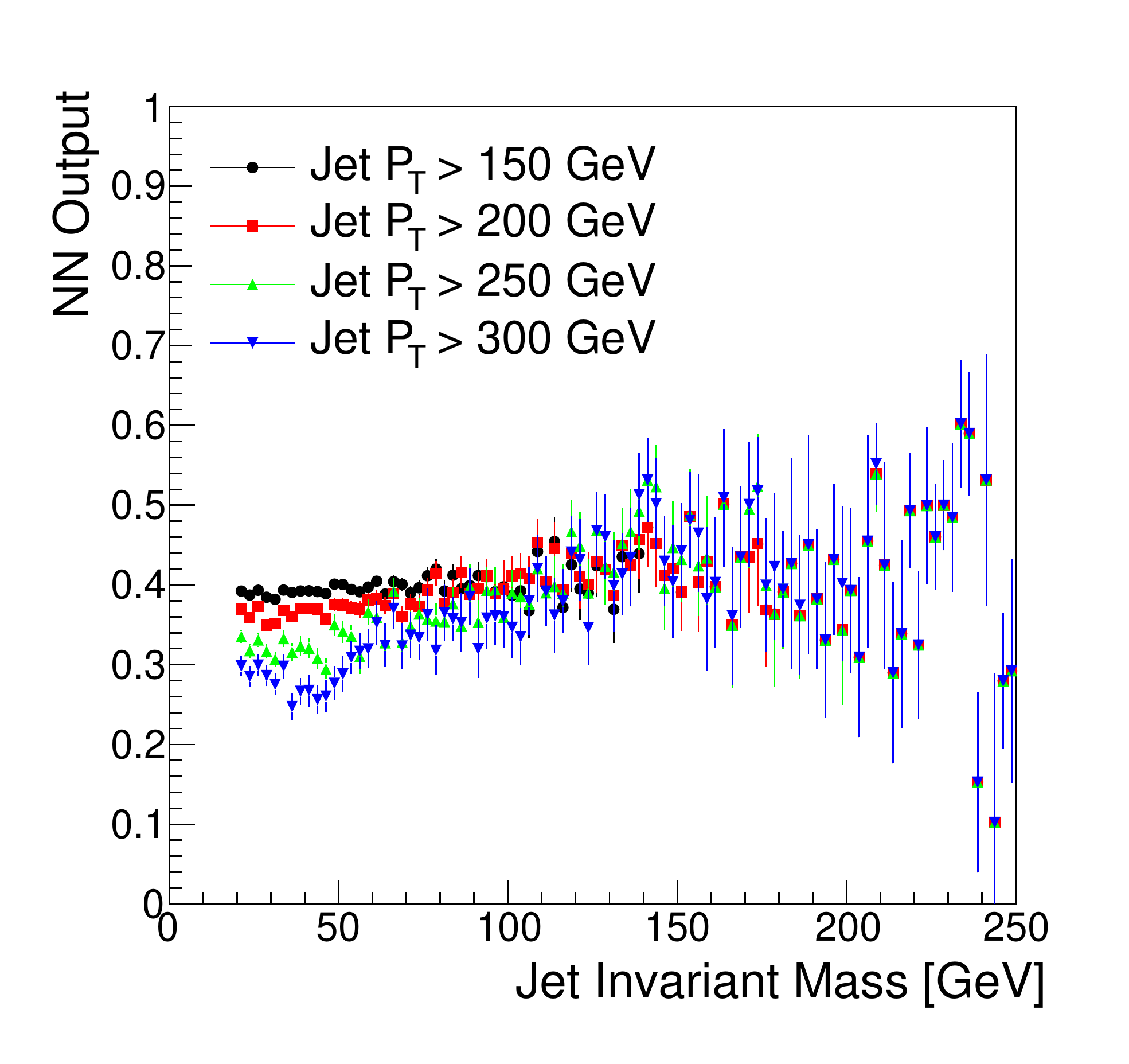}
\includegraphics[width=0.23\textwidth]{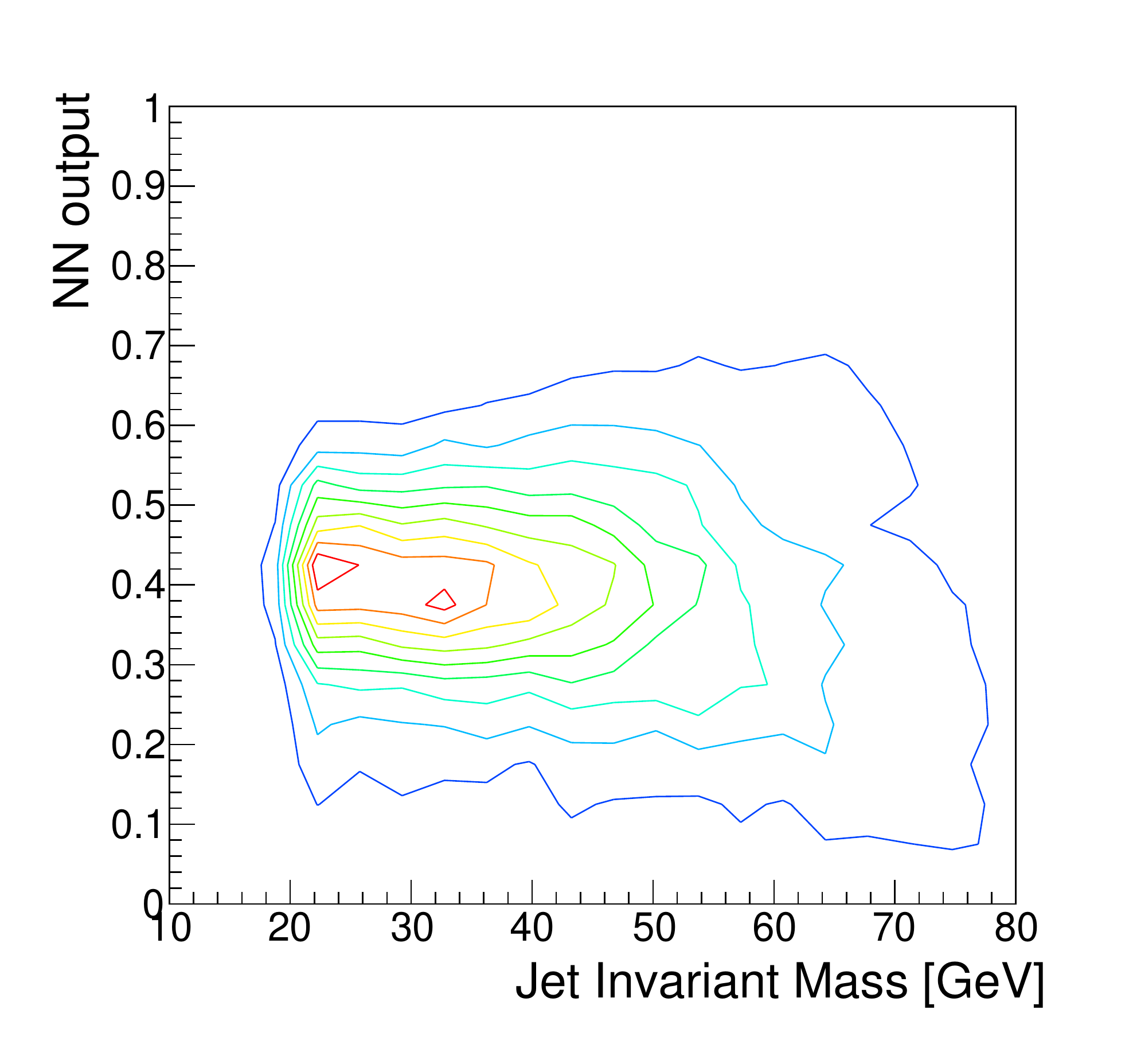}

\caption{ Top left, relationship between jet mass and neural network output in background events for a network trained to optimize classification compared to an adversarial network trained to optimize classification while minimizing dependence on jet mass.  Top right, relationship between jet mass and jet substructure variable $\tau_{21}$ and the DDT-modified $\tau'_{21}$ and $\tau''_{21}$ which attempt to minimize dependence on jet mass. Bottom left, profile of neural network output versus jet mass for the adversarial trained network with varying jet $p_{\textrm{T}}$ thresholds. Bottom right, contour plot of neural network output versus jet mass in background events for the adversarially-trained network. The signal sample used in training has $m_{Z'}=100$ GeV; generalization to other masses is shown in Sec.~\ref{sec:param}. }
\label{fig:profs}
\end{center}
\end{figure}

\begin{figure}[h!]
\begin{center}
\includegraphics[width=0.23\textwidth]{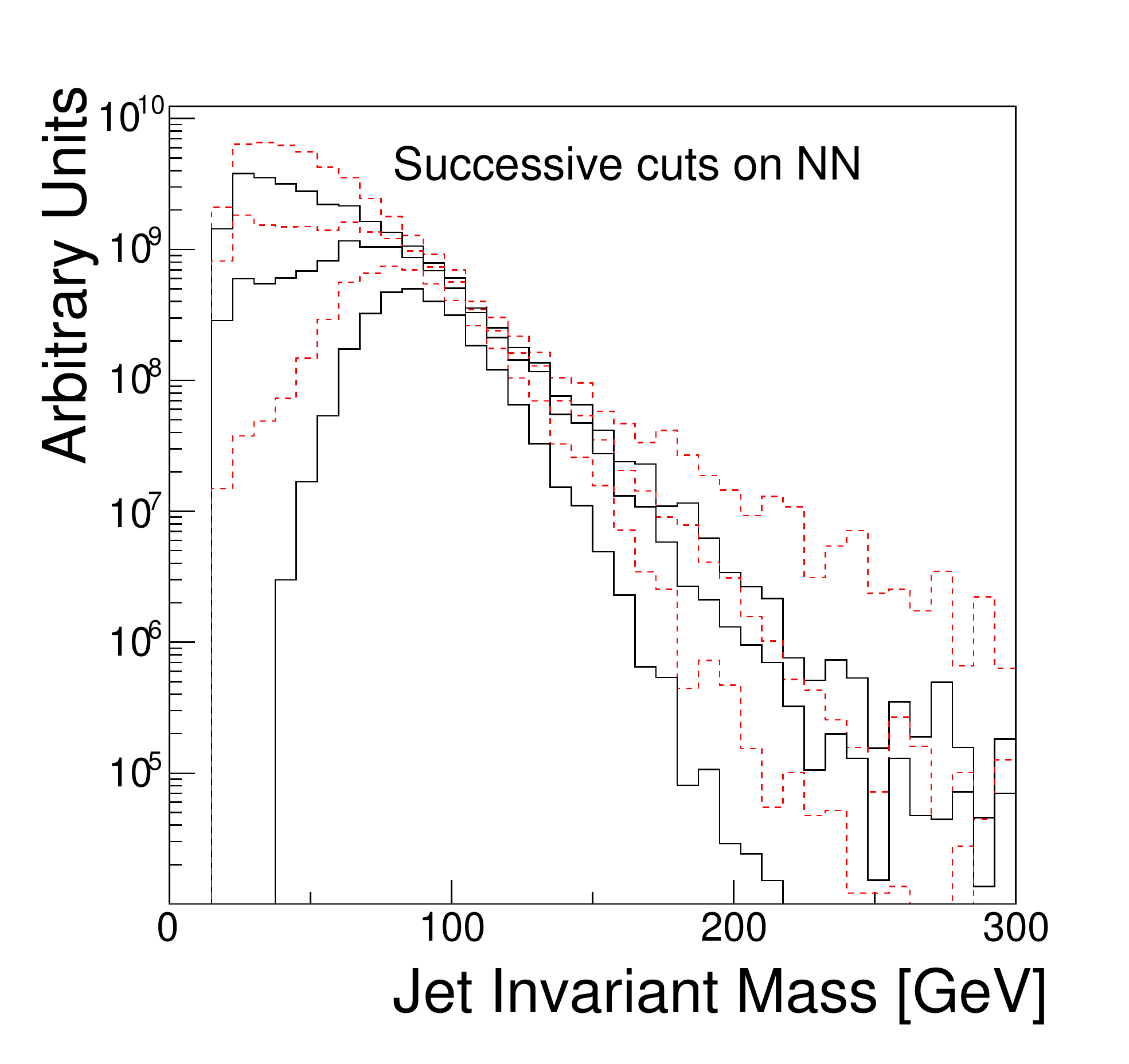}
\includegraphics[width=0.23\textwidth]{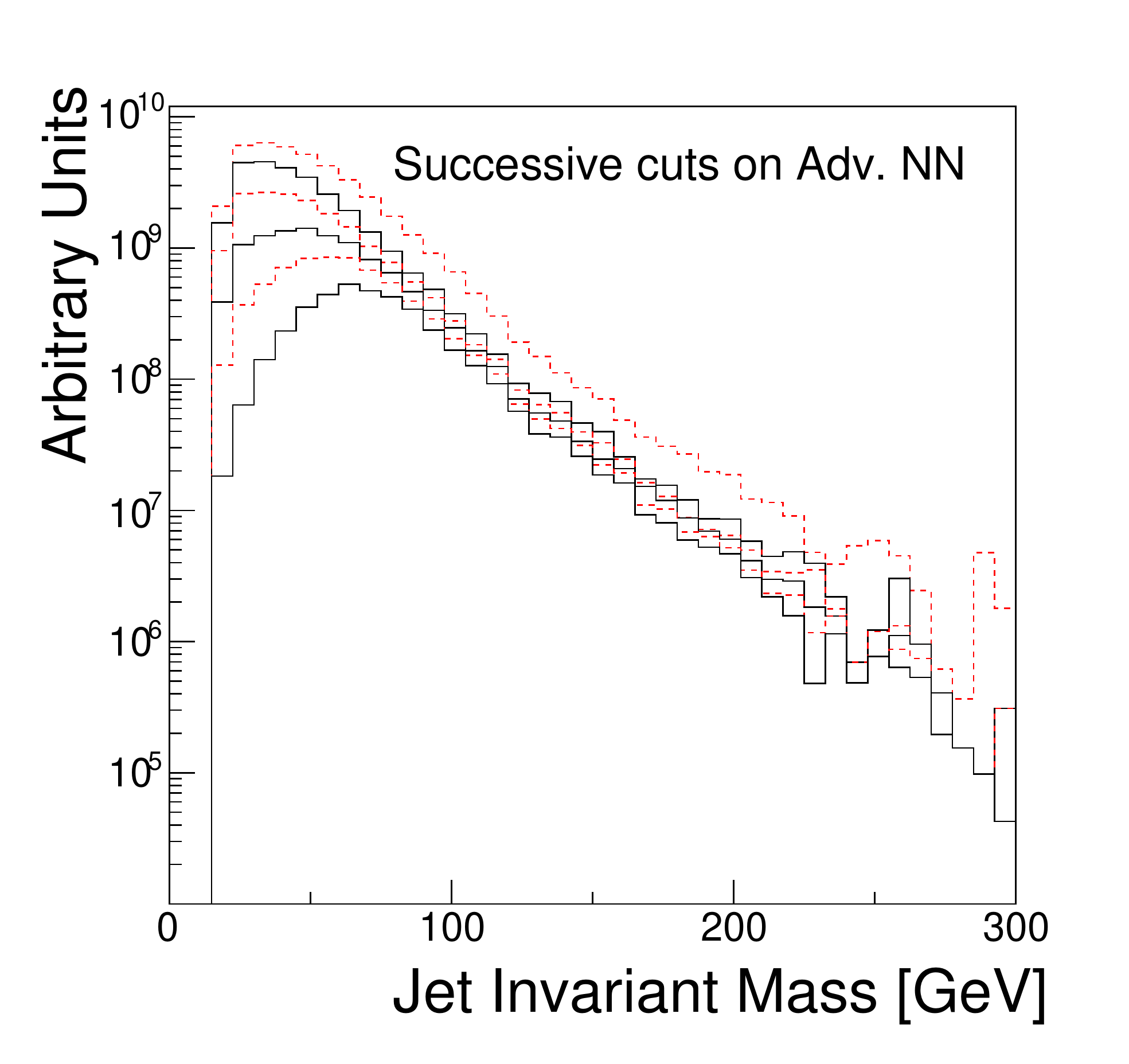}
\includegraphics[width=0.23\textwidth]{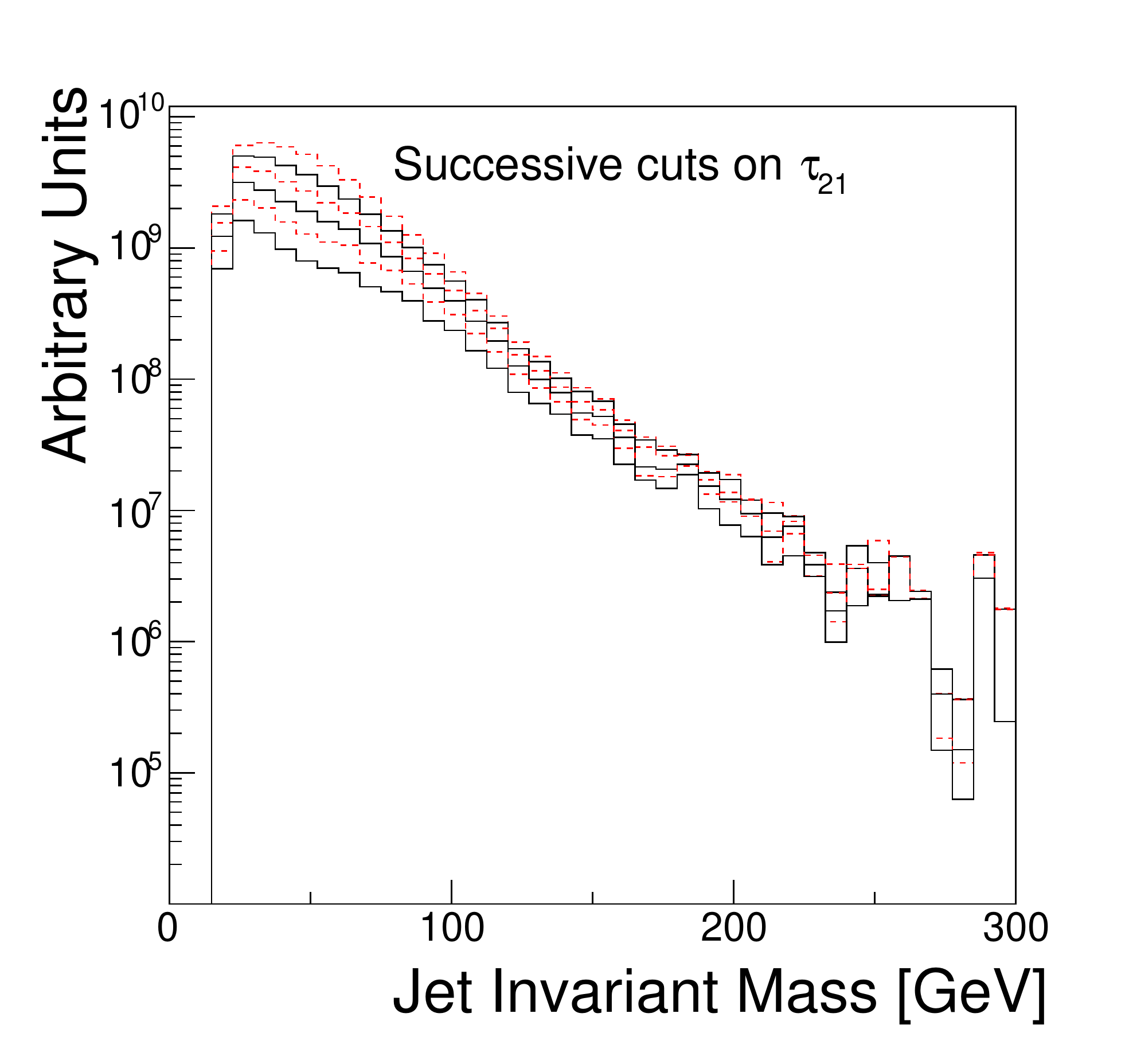}
\includegraphics[width=0.23\textwidth]{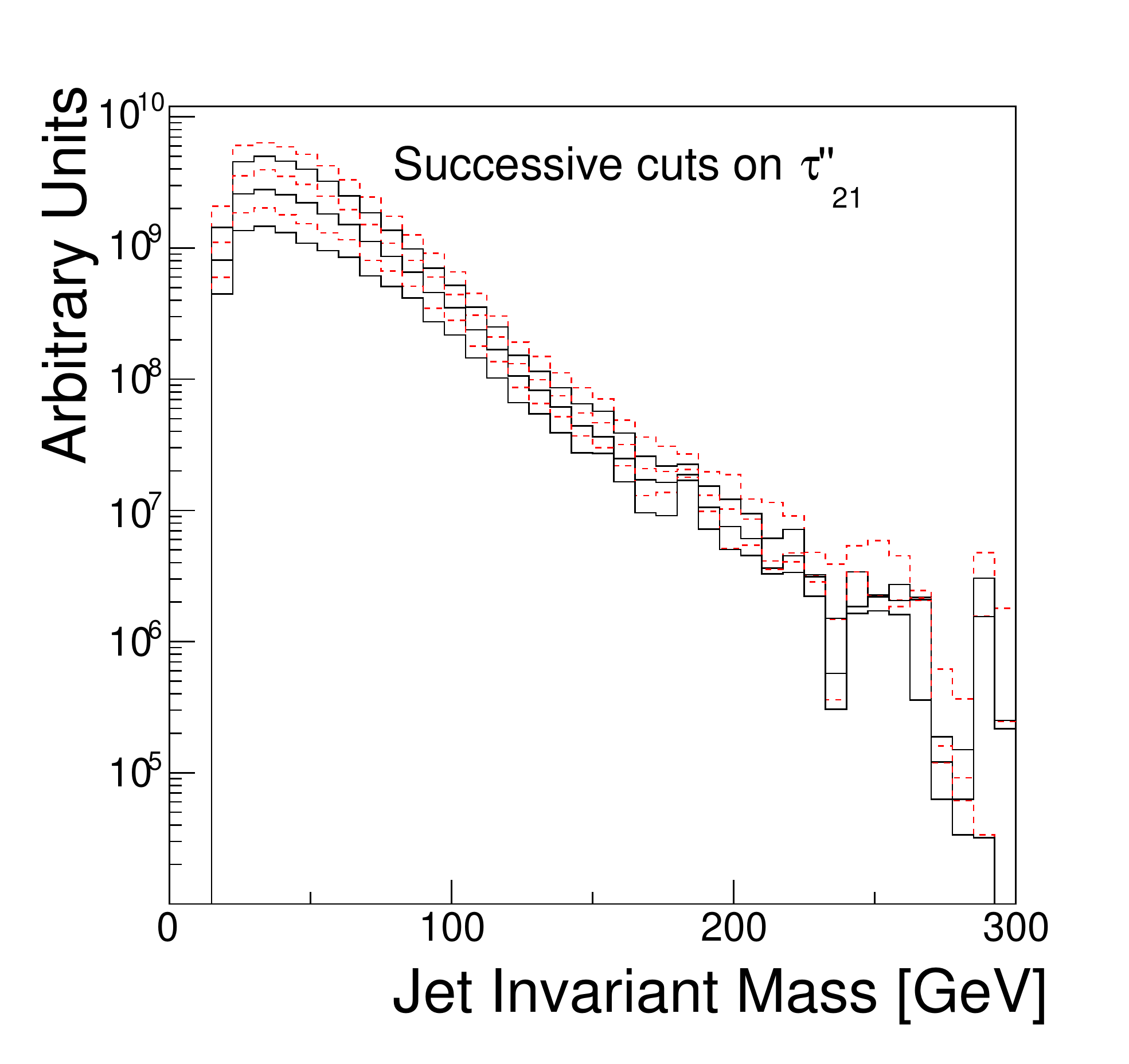}
\caption{ Jet mass distributions for background events with successively stricter requirements on different substructure discrimination strategies, giving signal efficiencies of $\varepsilon_{\textrm{sig}}=50,60,70,80,90,100$\%. Shown are the impact of threshold requirements on a neural network output  trained to optimize classification, an adversarial network which attempts to minimize depenence on jet mass,  $\tau_{21}$ and $\tau''_{21}$.}
\label{fig:slices}
\end{center}
\end{figure}

The focus of this study, however, is to look beyond the pure discriminatory power of these tools and study their effect on the jet mass spectrum. In Fig.~\ref{fig:profs}, it can be seen that the adversarial network output for background events has a profile which is largely independent of jet mass, while the classifying network is strongly dependent on jet mass.  Similarly, $\tau'_{21}$ and $\tau''_{21}$ have a lessened dependence on jet mass, compared to $\tau_{21}$. Figure~\ref{fig:slices} shows the effect on the jet mass distribution of successively stricter requirements on these variables. Note that the adversarial network's dependence on jet mass is diminished, but not eliminated, as can be seen in the contour plot of Fig.~\ref{fig:profs}. This is a reflection of the trade-off inherent in balancing classification power with jet mass dependence.
 
In Fig.~\ref{fig:profs}, we also show the profile of the neural network output versus jet mass, for various thresholds on the jet $p_{\textrm{T}}$, which shows some small $p_{\textrm{T}}$-dependent effects, but no large features.  As an alternative strategy, we trained a network using an adversarial strategy with respect to log($m/p_{\textrm{T}}$), which more closely mimics the approach used in Ref.~\cite{Dolen:2016kst}; the training succeeded in finding a network with a flat response in log($m/p_{\textrm{T}}$), but the distortion in jet mass was much more significant. In principle, it is possible to use the adversary to enforce a two-dimensional decorrelation, but since the $p_{\textrm{T}}$-dependence is not severe here, we leave this for future study.

\section{Statistical Interpretation}
\label{sec:stats}

The ability to discriminate jets due the hadronic decay of a boosted object from those due to a quark or gluon is an important feature of a jet substructure tagging tool, but as discussed above it is not the only requirement.  Due to the necessity of accurately modeling the background, it is desirable that the jet tagger avoid distortion of the background distribution.  Simpler background shapes are especially preferred because they allow for robust estimates that are constrained by the sidebands; backgrounds that can be modeled with fewer parameters and inflections avoid degeneracy with signal features, such as a peak.

Fig.~\ref{fig:profs} shows qualitatively that the adversarial network's response is not strongly dependent on jet mass. But a quantitative assessment is more difficult. Mass-independence is not in itself the goal; instead, we seek reduced dependence on knowledge of the background shape and reduced sensitivity to the systematic uncertainties that tend to dilute the statistical significance of a discovery.

However, our lack of knowledge of the true background model in general also makes it non-trivial to rigorously define and estimate the background uncertainty. In practice, experimentalists use an assumed functional form, with parameters constrained by background-dominated sidebands to predict the background in the signal region.
These assumptions may be validated by examining control regions in which the signal is not present, and the background processes are expected to exhibit physically similar properties.
For example, the tagger selection may be inverted to yield a sample with high background purity which may be used as a template.
If the tagger selection induces a distortion of the spectrum, these techniques are ineffective.
Moreover, when tagger-induced distortion depletes data from the sidebands (as is typically the case), any background model becomes more difficult to constrain.
To demonstrate these effects on the overall statistical performance of a search, we construct a simplified statistical test which has the desired behavior of penalizing discriminators which yield excessive distortion of the background shape.

A threshold is placed on the discriminator output, after which a likelihood fit is performed, binned in the distribution of reconstructed large-radius jet masses using signal and background templates from simulated samples\footnote{In principle, the most powerful approach is a likelihood directly on the output of the discriminator, but this requires a valid model of the background, which is lacking in this case.}. An uncertainty on the rate of the background is included in order to model our lack of knowledge of the background.   We calculate expected discovery significance using a profile likelihood ratio~\cite{Cowan:2010js} with the CLs technique~\cite{Read:2002hq,Junk:1999kv}, marginalizing over the unknown background rate.

Though the background shape is fixed via the template, the uncertainty on the rate provides the statistical behavior we seek. Specifically, if the uncertainty in the rate of the background is large enough, then the discovery significance is sensitive also to the shape of the background distribution as follows. In the case that the background is fairly flat, there are background-dominated sidebands which can constrain the rate uncertainty. In the opposite case that the background is distorted to mimic the signal, these sideband constraints have reduced power, and the signal and background are more difficult to distinguish statistically.
Hence, the presence of rate uncertainties penalizes a solution which distorts the background spectrum as desired.
Although this simple approach likely underestimates the true impact of more realistic systematics, it is sufficient to illustrate the effect on sensitivity.
In the following, we take for the small (large)-uncertainty case a relative uncertainty of 5\% (50\%) on the overall background rate.

\begin{figure}[h!]
\begin{center}
\includegraphics[width=0.23\textwidth]{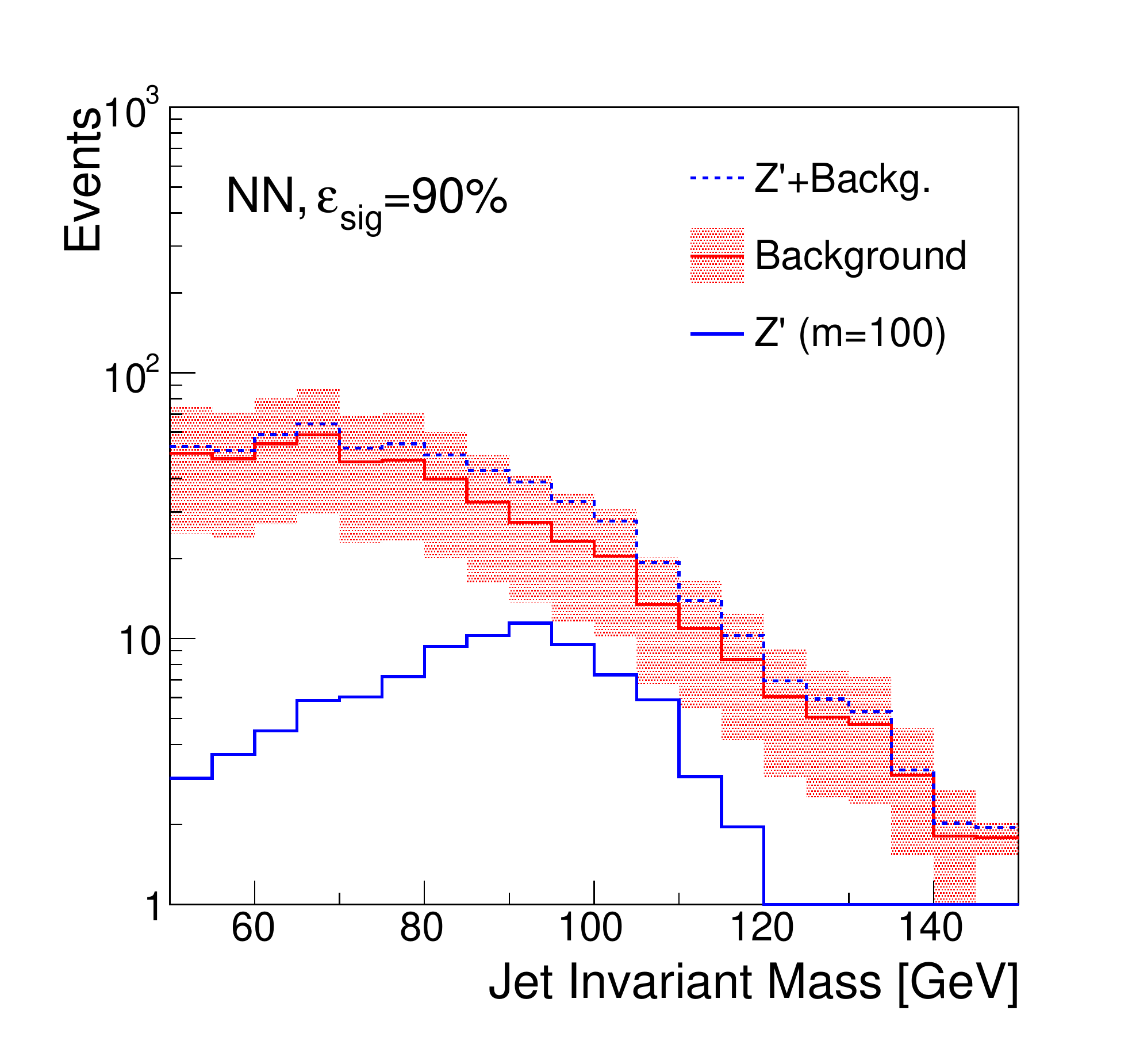}
\includegraphics[width=0.23\textwidth]{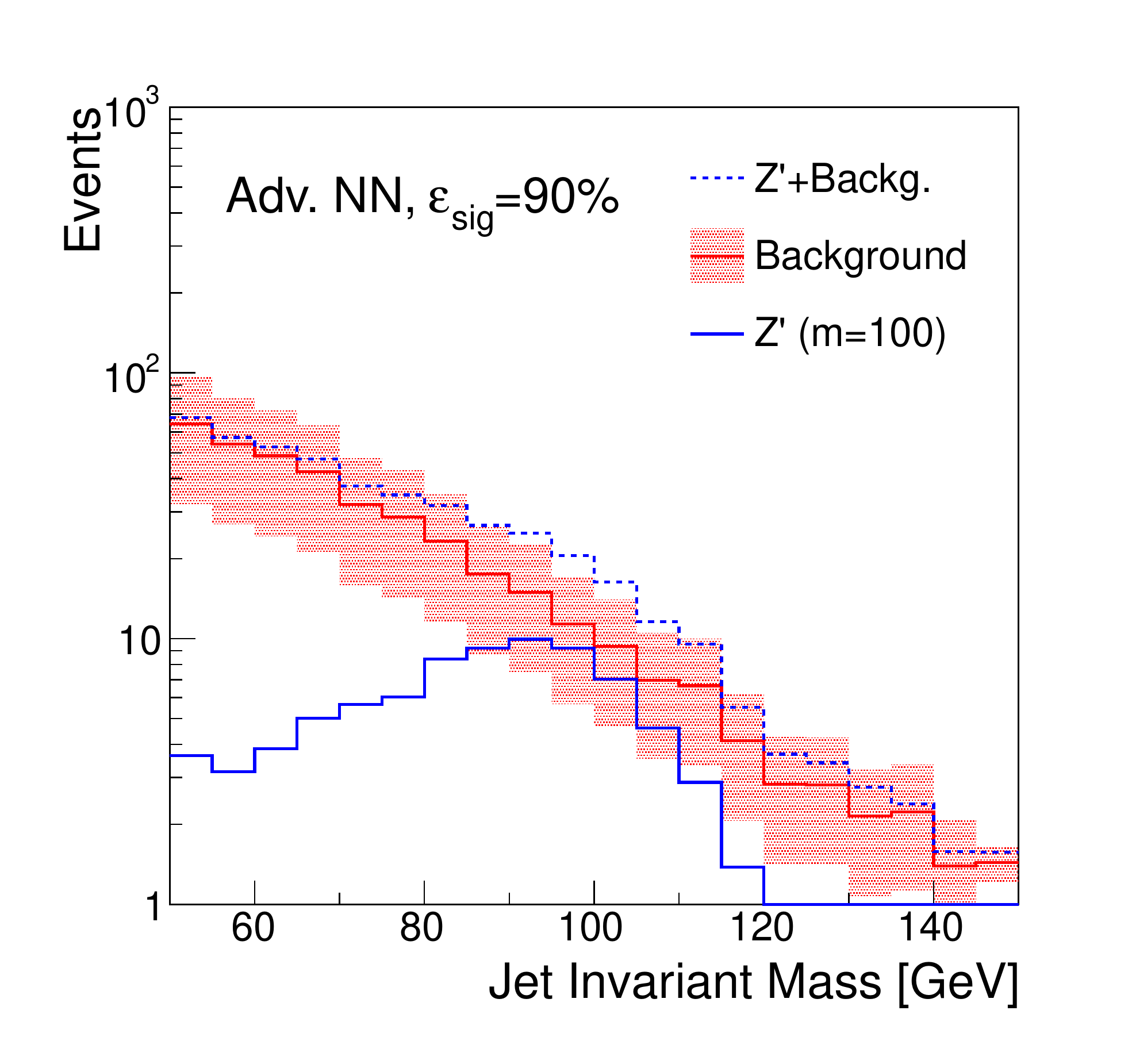}
\includegraphics[width=0.23\textwidth]{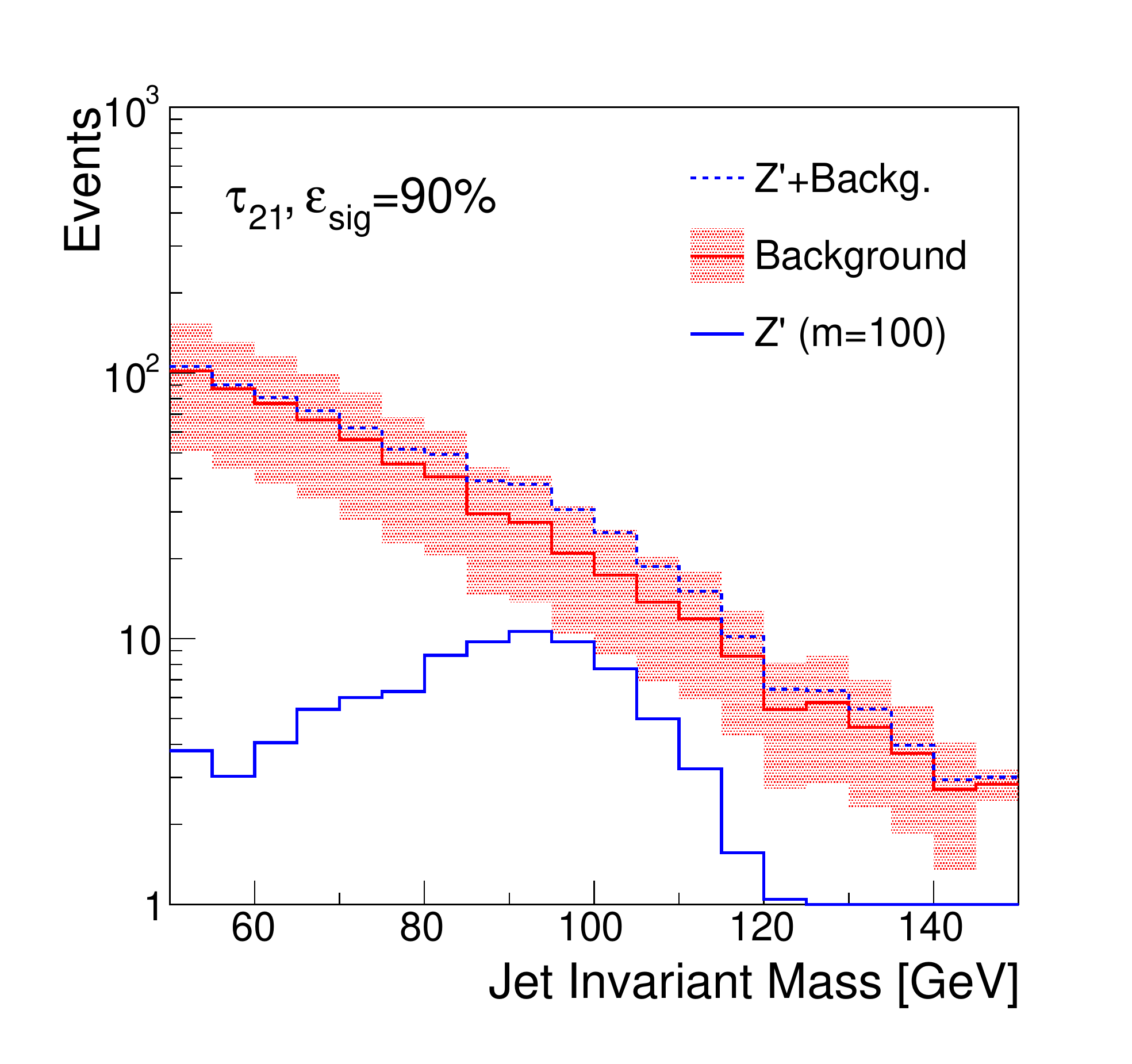}
\includegraphics[width=0.23\textwidth]{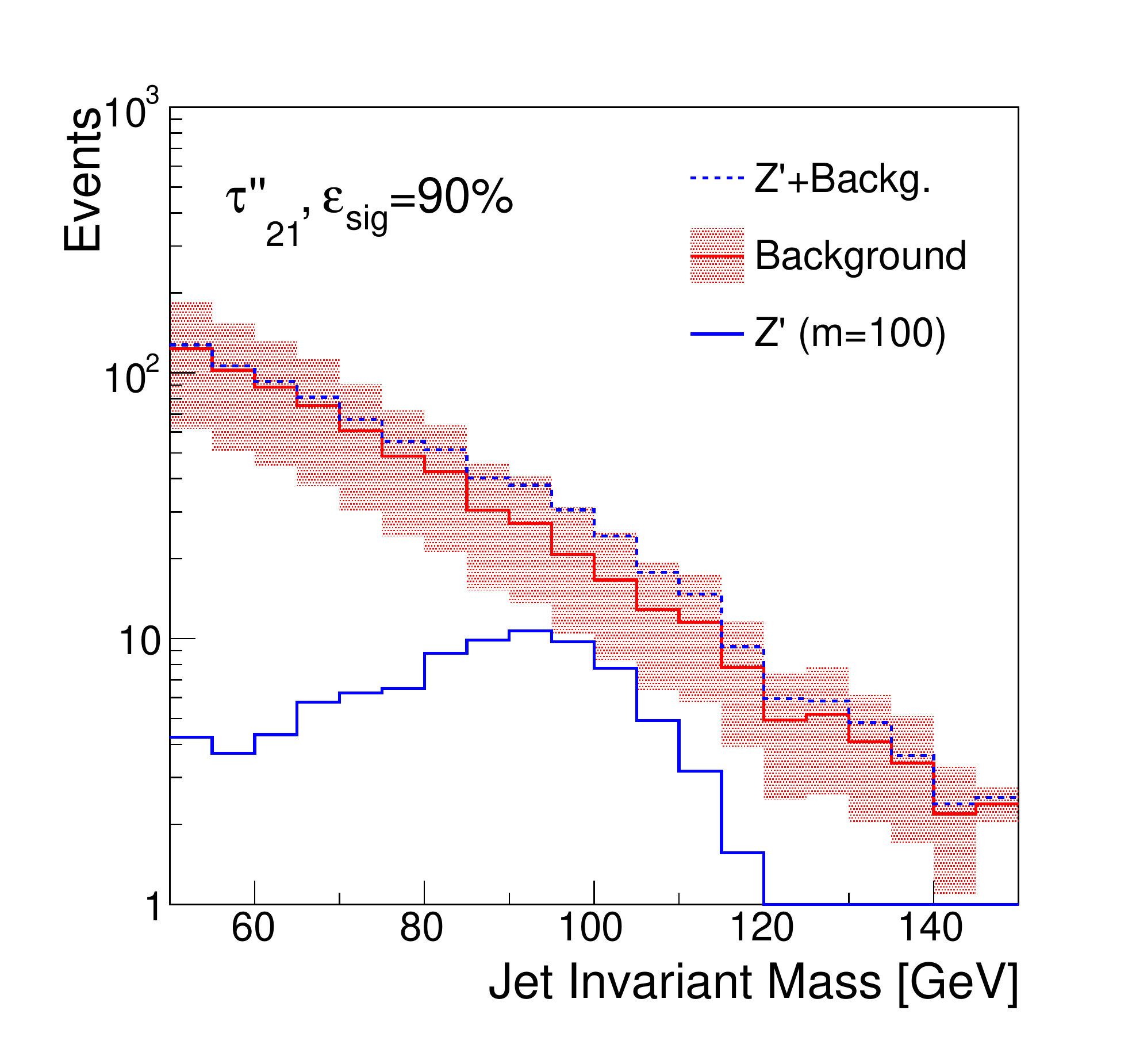}
\caption{Distributions of jet mass after selection with signal efficiency of 90\% using the NN classifier, the adversarial network, $\tau_{21}$ or $\tau''_{21}$. Background distributions are shown with 50\% uncertainty.}
\label{fig:hist_90}
\end{center}
\end{figure}

Examples of the final jet mass distribution are shown in Figs.~\ref{fig:hist_90} and~\ref{fig:hist_50} for thresholds on the discriminants which result in signal efficiency of 90\% and 50\% respectively.

\section{Results}

The discovery significance is measured for varying thresholds on the discriminator outputs. While all of the discriminators exhibit some degree of classification power, this study explores the question of whether they provide additional discovery significance.

\begin{figure}[h!]
\begin{center}
\includegraphics[width=0.23\textwidth]{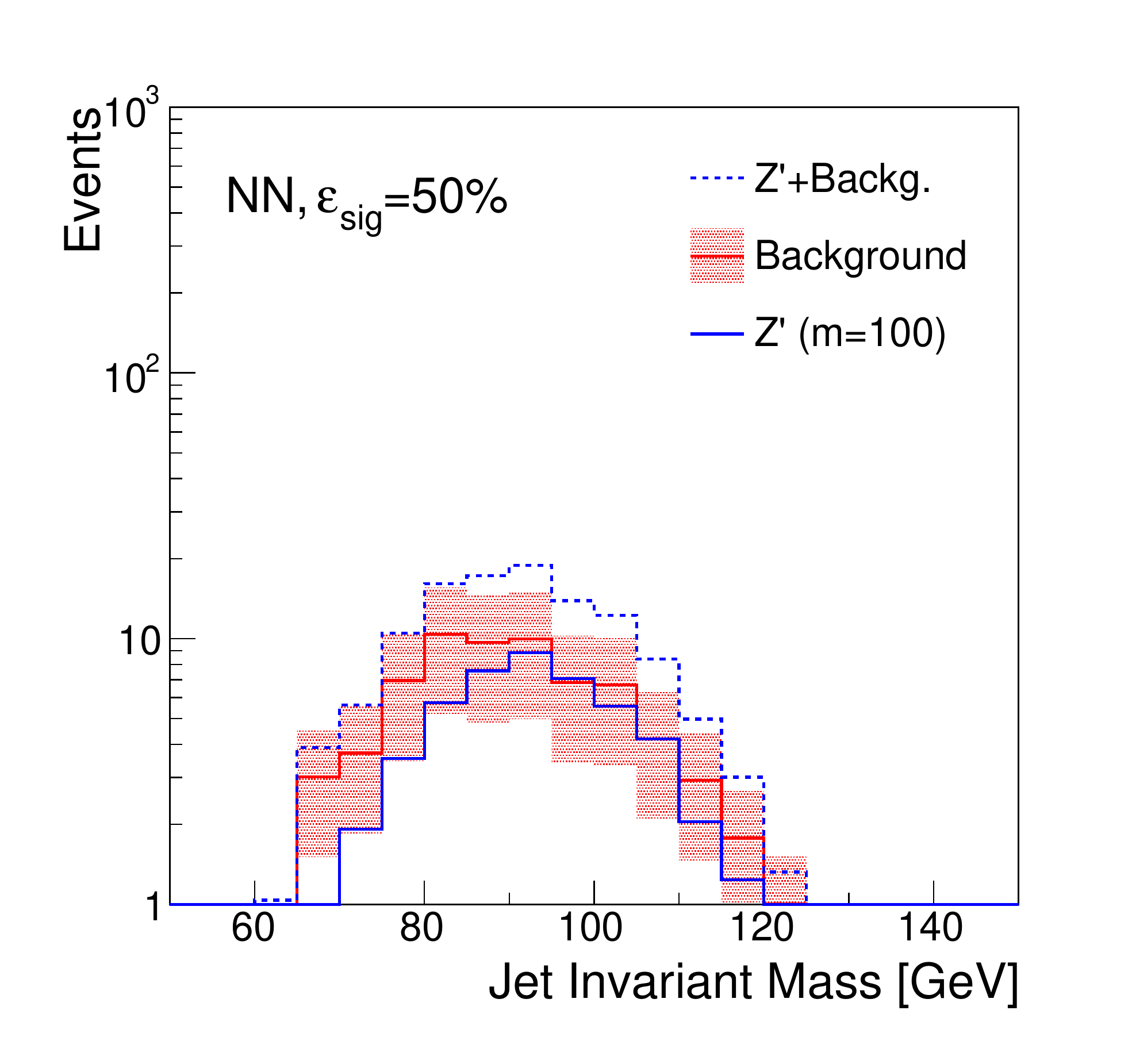}
\includegraphics[width=0.23\textwidth]{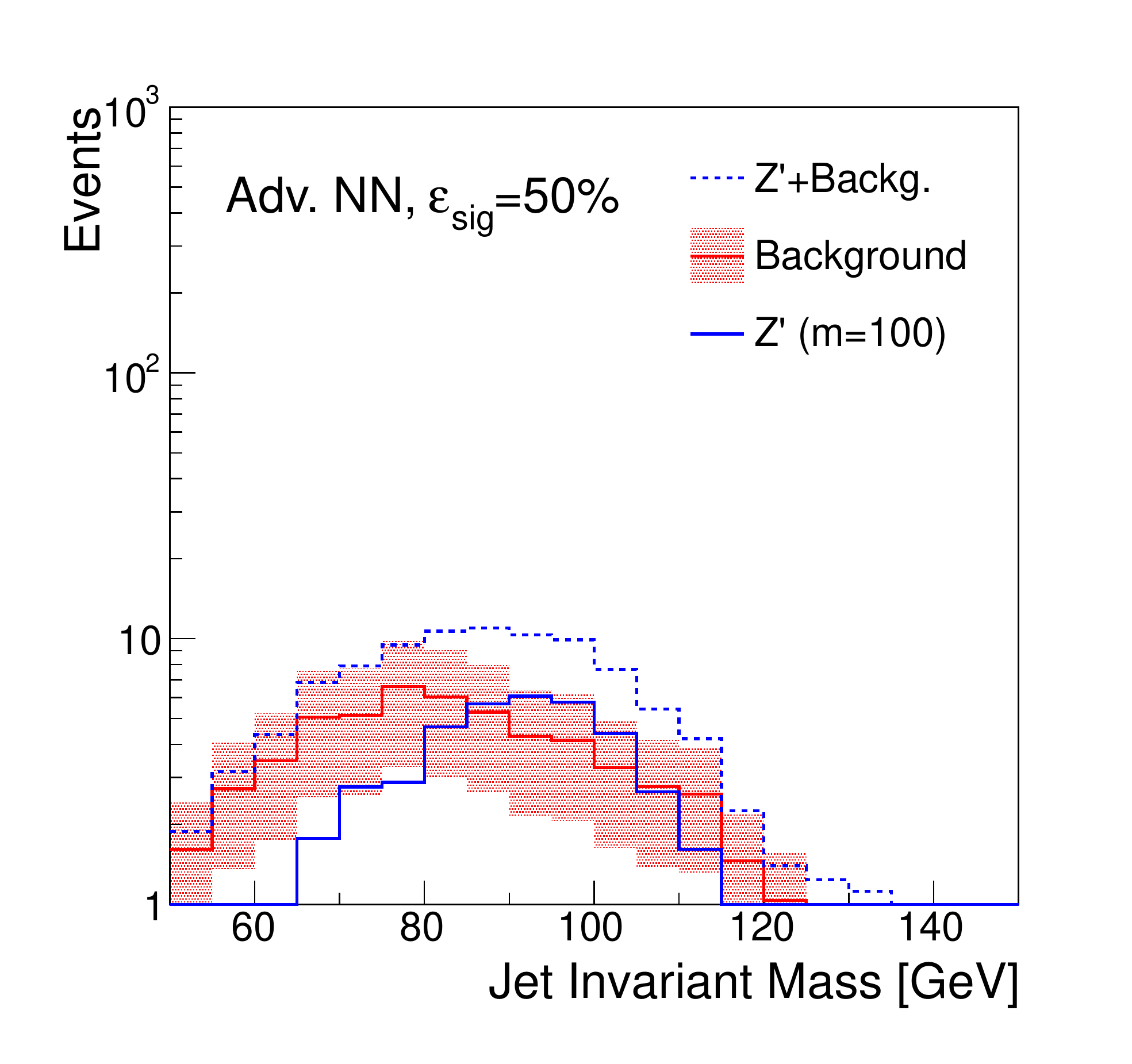}
\includegraphics[width=0.23\textwidth]{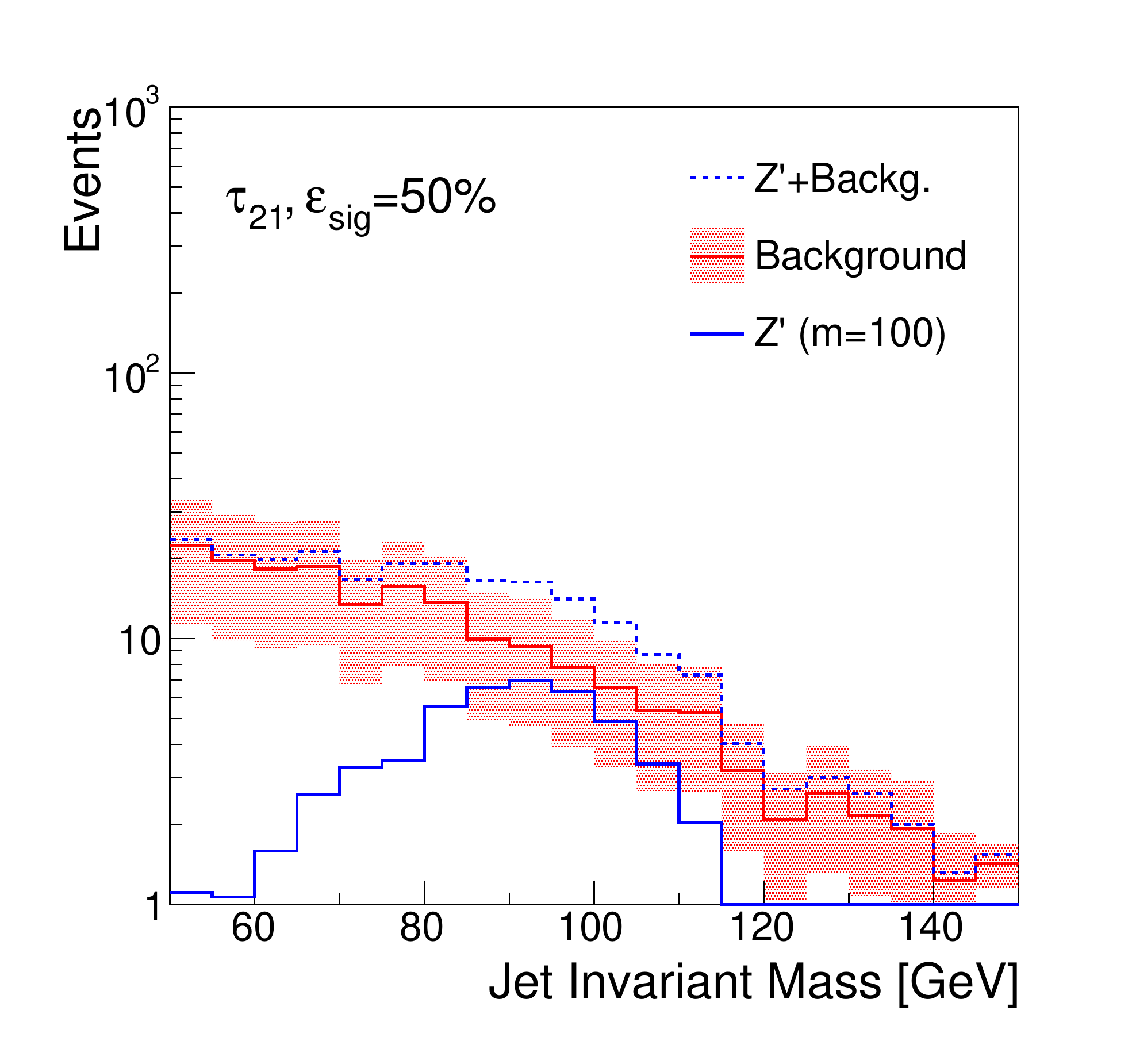}
\includegraphics[width=0.23\textwidth]{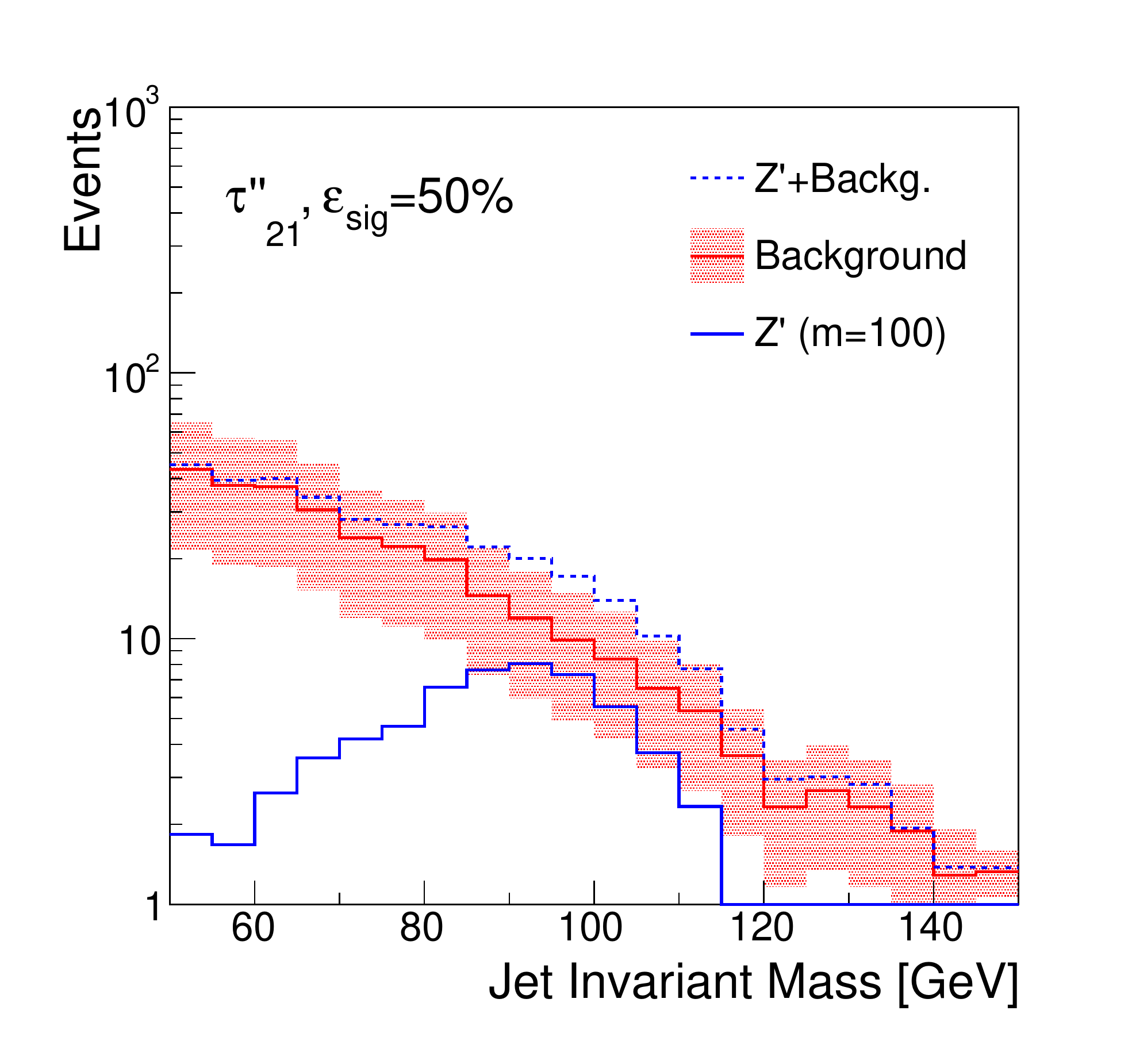}
\caption{Distributions of jet mass after selection with signal efficiency of 50\% using the NN classifier, the adversarial network, $\tau_{21}$ or $\tau''_{21}$.  Background distributions are shown with 50\% uncertainty.}
\label{fig:hist_50}
\end{center}
\end{figure}

\begin{figure}[h!]
\begin{center}
\includegraphics[width=0.45\textwidth]{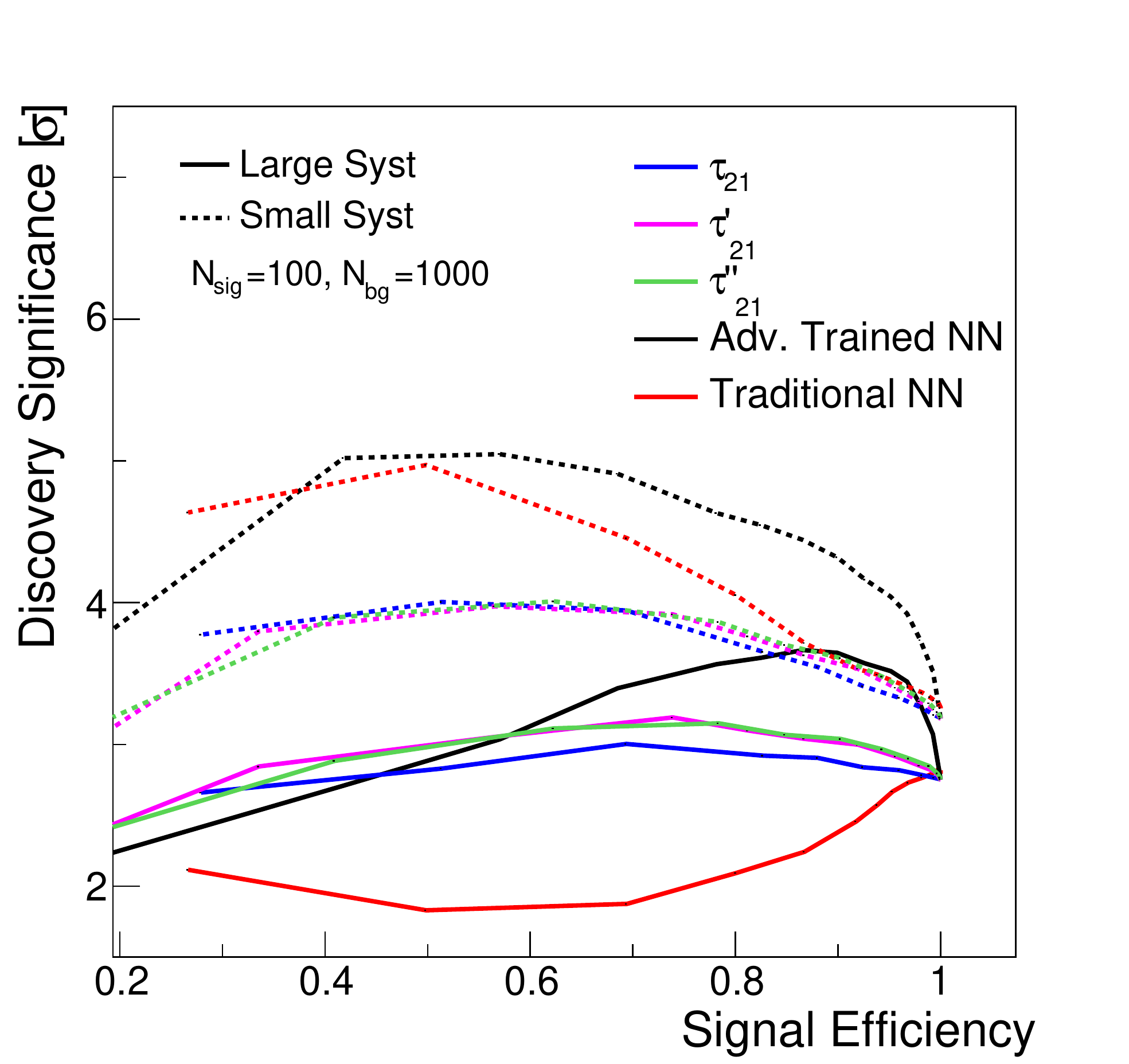}
\caption{ Statistical significance of a hypothetical signal for varying thresholds on the outputs of networks trained to optimize classification compared to adversarial networks trained to optimize classification while minimizing impact on jet mass. Shown are two scenarios, in which the uncertainty on the background level is negligible or large, both with $N_{\textrm{sig}}=100, N_{\textrm{bg}}=1000$.  }
\label{fig:signif}
\end{center}
\end{figure}

Figure~\ref{fig:signif} shows the discovery significance as a function of the signal efficiency of the discriminator threshold, for two choices of background uncertainty. In the case of the small uncertainty (5\% relative), applying a tighter threshold on the discriminator improves the discovery significance, despite lowering the signal efficiency, due to the heightened background suppression.  Even at fairly low signal efficiencies of 50\%, where the background is sculpted to look like the signal (see Fig.~\ref{fig:hist_50}), the discovery significance is improved. This is as expected; if the background rate and shape are well known, then the lack of constraining sidebands is not detrimental.

For the case of the larger background rate uncertainty, thresholds on $\tau_{21}$ provide a smaller boost to the significance. The large relative uncertainty on the background will penalize configurations in which the background is sculpted to resemble the signal, preventing the data from constraining the background rate in the sidebands.  Thresholds on $\tau'_{21}$ and $\tau''_{21}$ are slightly stronger, as expected, due to their decreased correlation with jet mass.  Thresholds on the output of the classifier network, which has the strongest discrimination power, only weakens the discovery significance, due to the background mass distortion. However, the adversarial network is still capable of powerful discrimination which improves the discovery power at high signal efficiency, around 90\%. Table~\ref{tab:signif} shows the maximal discovery significance for each case. The qualitative results persist for other signal-to-background ratios.

\begin{table}
\caption{ Signal and background efficiencies at maximal discovery significance at $m_{Z'}=100$ GeV for each method and for scenarios of large (50\%) or small (5\%) relative systematic uncertainty on the background rate.}
\label{tab:signif}
\begin{tabular}{lrrr}
\hline
\hline
Method &\ \ Signal \ \  & Background & Discovery \\  
 &\ \  Eff.\ \  &  Eff. & Signif. ($\sigma$) \\  
\hline
\hline
\multicolumn{4}{c}{{\emph{5\% background uncertainty}}}\\
\hline
Adv. Trained NN	&0.44 & 0.06	&5.05\\  
Traditional NN	&0.39 & 0.03	&4.97\\ 
$\tau_{21}$	&0.44 & 0.19	&4.00\\ 
$\tau'_{21}$	&0.50 & 0.29	&3.97\\ 
$\tau''_{21}$   &0.52 & 0.26   &4.01\\ 
\hline
\multicolumn{4}{c}{\emph{50\% background uncertainty}}\\
\hline
Adv. Trained NN	&0.82 & 0.48	&3.67\\ 
Traditional NN	&1.00 & 1.00	&2.82\\ 
$\tau_{21}$	&0.60 & 0.32	&3.00\\ 
$\tau'_{21}$	&0.70 & 0.50	&3.19\\ 
$\tau''_{21}$   &0.70 & 0.45   &3.15\\
\hline
\hline
\end{tabular}
\end{table}

\section{Parameterized Neural Networks}
\label{sec:param}

The studies above demonstrate the application for the case of a single example value of the hypothetical $Z'$ mass. In this section, we show that the same approach can be generalized to solve a set of closely related problems, jet classification for different $Z'$ masses, using a single neural network parameterized in $m_{Z'}$. 

\begin{figure}[h!]
\begin{center}
\includegraphics[width=0.45\textwidth]{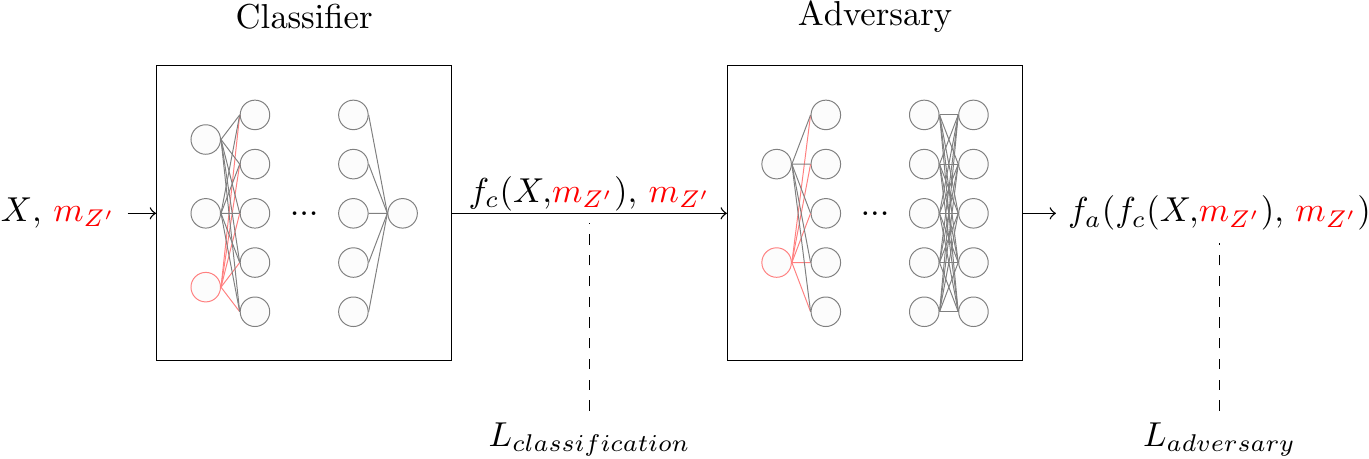}
\caption{ Architecture of the neural networks in the parameterized adversarial training strategy. The classifying network distinguishes signal from background using the eleven variables described in the text ($X$) plus $m_{Z'}$. The classifying network output is then a function of $m_{Z'}$.  The adversarial network attempts to predict the invariant mass using the output of the classifier, $f_{c}(X,m_{Z'})$ as well as $m_{Z'}$. }
\label{fig:network_param}
\end{center}
\end{figure}

These parameterized neural networks~\cite{Baldi:2016fzo} address a common problem in physics: solving a classification task multiple times for different values of an unknown latent variable, like $m_{Z'}$. Simulations used to train jet classifiers are generally performed for a small set of fixed $Z'$ mass values. In the traditional approach, a separate neural network classifier is trained for each $Z'$ mass value. However, by treating $m_{Z'}$ as just another input feature, a single parameterized neural network can learn to solve the related classification tasks all at once (Fig. \ref{fig:network_param}). Furthermore, the classifier can interpolate to other values of $m_{Z'}$ if the function is smooth. 

For this experiment, some hyperparameters were tuned to this more complex task. The classifier had three hidden layers of 300 $\tanh$ nodes, with a learning rate of $10^-4$, a momentum of $0.95$, and an L2 weight decay factor of $10^-3$ in each layer. The adversary consisted of two hidden layers of 100 $\tanh$ nodes each, with a learning rate of $10^-2$, a momentum of $0.95$, and an L2 weight decay factor of $10^-4$ in each layer. The parameter $\lambda$ was set to $10$.

\begin{figure}[h!]
\begin{center}
\includegraphics[width=0.45\textwidth]{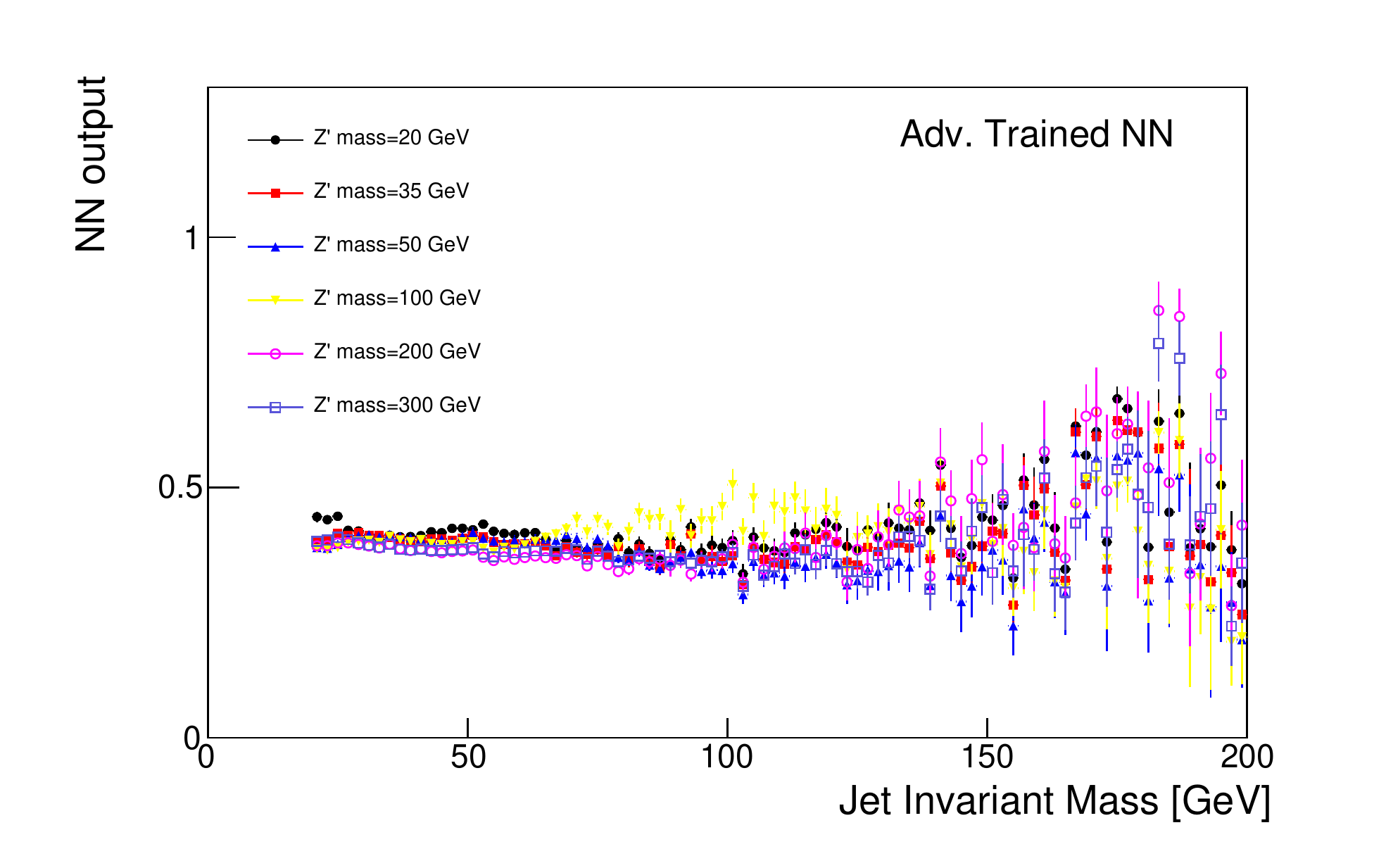}
\includegraphics[width=0.45\textwidth]{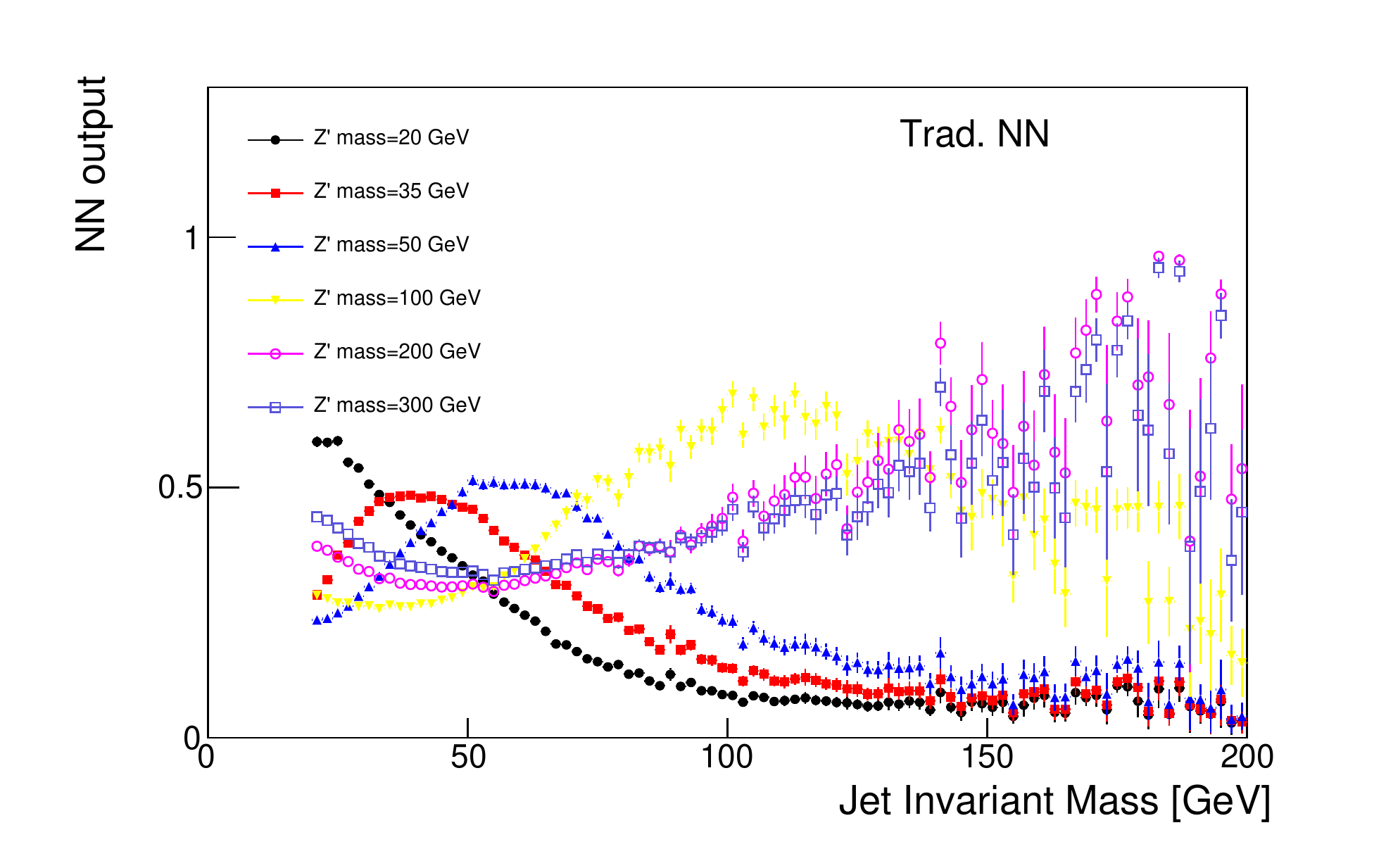}
\caption{
Profile of the paramterized NN responses to background versus jet mass, where the parameterized network was evaluated at different $Z'$ mass hypotheses. Top shows the response of the adversarially-trained classifier, which minimizes correlation with jet mass; bottom shows the response of a network trained in the traditional manner, to optimize classification accuracy.}
\label{fig:profparam}
\end{center}
\end{figure}

The adversary was also parameterized by including the $Z'$ mass as an input along with the classifier output. The resulting classifier predictions for background events are mostly independent of mass when conditioned on each theory mass (Fig.~\ref{fig:profparam}). Without this parameterization of the adversary, the marginalized classifier predictions are independent of mass, but not the conditional classifier predictions.

\begin{figure}[h!]
\begin{center}
\includegraphics[width=0.4\textwidth]{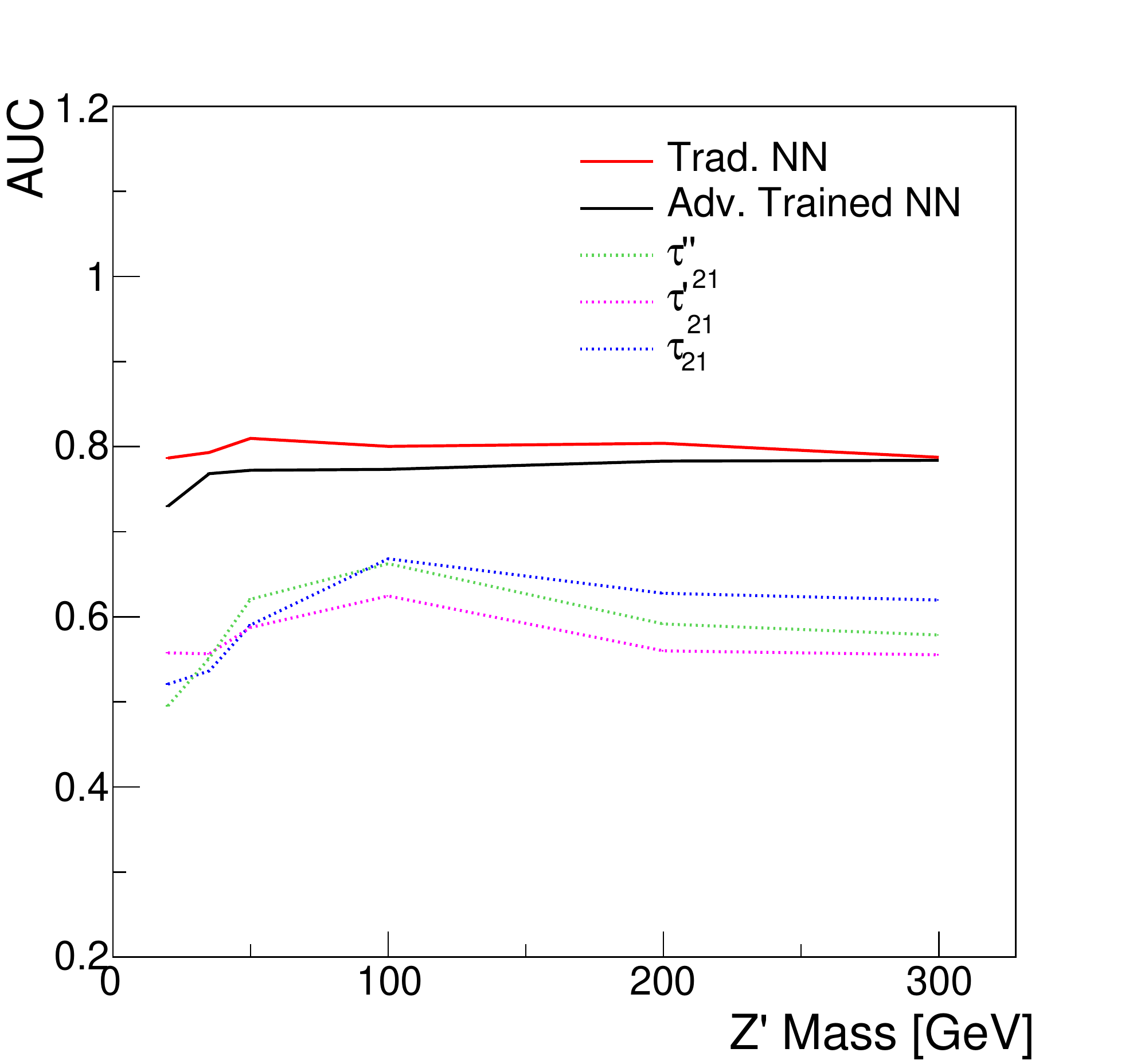}
\caption{ The AUC metric (Area Under the Curve) for NNs parameterized in $m_{Z'}$ and tested at several values (both traditional and adversarial training techniques), compared to the discrimination of the individual features $\tau_{21}$, $\tau'_{21}$, and $\tau''_{21}$.}
\label{fig:aucparam}
\end{center}
\end{figure}

As expected, the resulting classifier demonstrates better performance than the single input features $\tau_{21}$, $\tau'_{21}$ or $\tau''_{21}$ at all signal mass hypotheses tested (Fig.~\ref{fig:aucparam}).  As in the non-parameterized case, the traditional NN trained to maximize classification accuracy achieves the best separation.

\begin{figure}[h!]
\begin{center}
\includegraphics[width=0.45\textwidth]{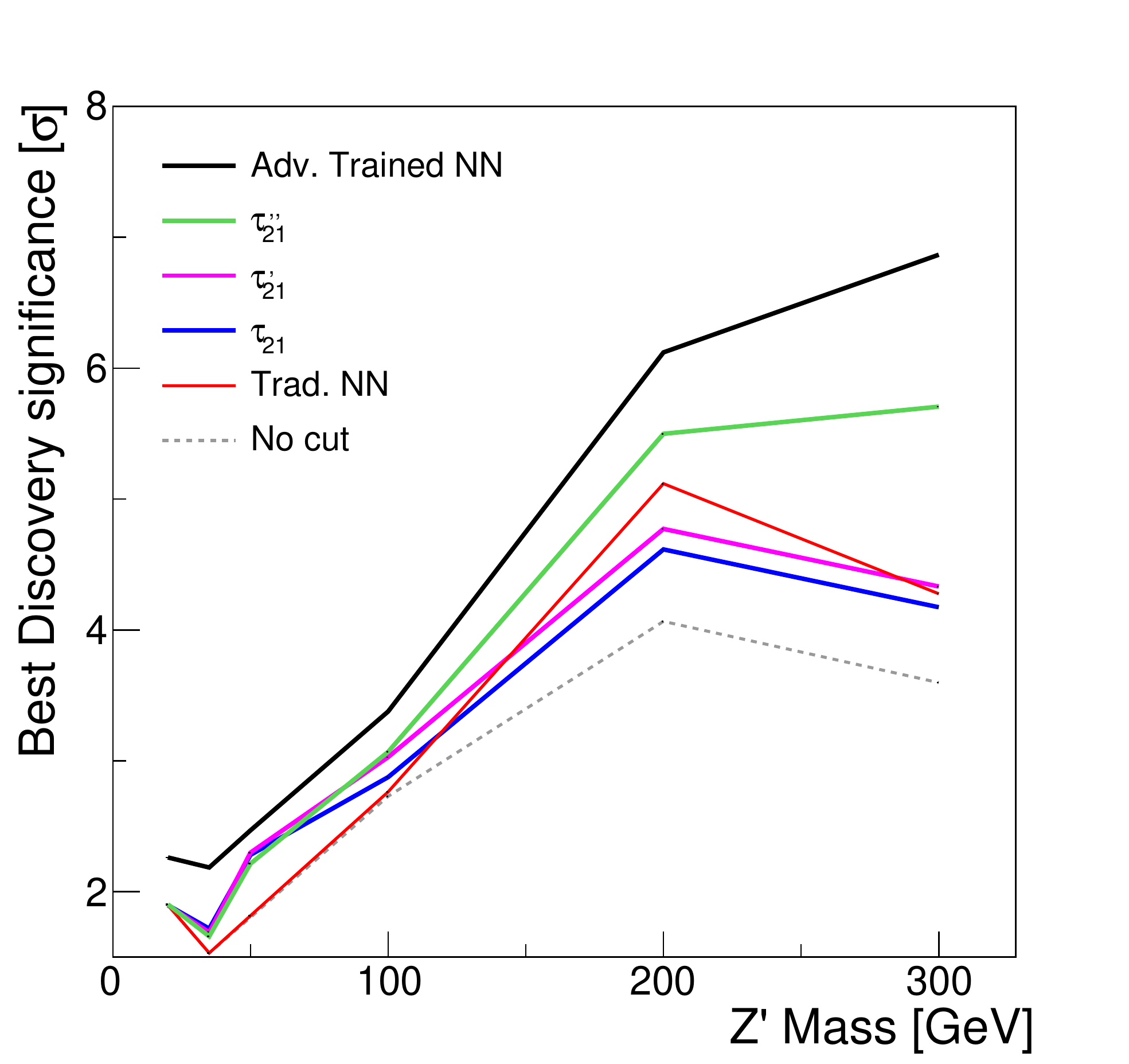}
\caption{ Discovery significance for a hypothetical signal after optimizing thresholds on the output of networks parameterized in $m_{Z'}$ trained with an adversarial or traditional approaches, compared to thresholds on  $\tau_{21}$, $\tau'_{21}$ and $\tau''_{21}$ or to placing no threshold.
Significance is evaluated for the case of 50\% background uncertainty.
}
\label{fig:sigparam}
\end{center}
\end{figure}

Moreover, the lack of background distortion by the adversarially-trained network preserves the ability to distinguish the background and signal mass distributions, leading to improved discovery significance;  see Fig.~\ref{fig:sigparam}. The statistical test is performed as for the previous case, fitting a binned likelihood on the jet mass distribution after applying a threshold on the discriminator output. As before, the improved separation of the traditional NN does not translate to improved discovery significance.

We note that while the performance shown here is evaluated on hypothesized mass values used for training, Ref.~\cite{Baldi:2016fzo} demonstrates this architecture is able to successfully interpolate to other values of $m_{Z'}$.

\section{Discussion}

We have demonstrated that an adversarial training strategy may yield a jet classification tagger which leverages the powerfully discriminating information obtained by combining several input features, while decorrelating its output from the variable of interest, the jet mass.
This allows the classifier to enhance signal to noise ratio while minimizing the tendency of the background distribution to morph into a shape which is degenerate with the observable signal.
When the background cannot be reliably predicted \emph{a priori}, as is often the case, it is important to be able to constrain its rate in sidebands surrounding the signal region.
Therefore, avoiding such degeneracy is critical to performing successful measurements.

We note that, from Fig.~\ref{fig:hist_50}, it is clear that applying sufficiently tight cuts to the adversarial classifier causes significant background morphing, particularly when compared to the $\tau_{21}$-based discriminants.
However, the solid lines of Fig.~\ref{fig:signif} illustrate the case where the background rate is uncertain and hence benefits from sideband constraints.
We see that the optimal significance is realized for the adversarial classifier at a relatively high signal efficiency of roughly 90\%, where the background morphing is quite limited (Fig.~\ref{fig:hist_90}).
Hence, the adversarial classifier achieves its goal of optimizing the tradeoff between correlation and discrimination power.

We also note that the decorrelation could potentially be improved.
The contour plot in Fig.~\ref{fig:profs} shows that while the average NN output is independent of mass, there is certainly still structure that results in the background sculpting still observed.
The residual $p_\mathrm{T}$ dependence could also be removed, possibly with a more sophisticated adversary that is trained to predict multiple variables simultaneously.
These improvements we leave for future work.

Finally, we extend the strategy to the case of a parameterized network wherein the NN classifier is trained to tag specific signal hypotheses, useful for scanning a range of theoretical parameter space with a search.
The resulting combined approach should be readily applicable to experimental measurements and searches, boosting their discovery significance or search sensitivity.

\section{Acknowledgements}

The authors acknowledge useful conversations with Kyle Cranmer, Jesse Thaler, Kevin Bauer, and Dan Guest,  helpful comments from Sal Rappoccio, Derek Soeder and Michela Paganini, and are grateful to the Aspen Center for Physics, where much useful discussion occured.

\bibliography{adv}

\end{document}